\pdfoutput=1
\documentclass[letterpaper,twocolumn,10pt]{article}
\PassOptionsToPackage{hyphens}{url}
\usepackage{usenix2019_v3}

\usepackage{tikz}
\usetikzlibrary{positioning,arrows.meta,fit}
\usepackage{amsmath}

\usepackage{booktabs}
\usepackage{array}
\newcolumntype{L}[1]{>{\raggedright\arraybackslash}p{#1}}

\newcommand{\gap}[1]{\textcolor{red}{\textbf{??}}}

\usepackage{textcomp}
\usepackage{listings}
\usepackage[tableposition=top]{caption}

\newcommand{\BS}{\textbackslash}

\usepackage{pifont}
\usepackage[inline]{enumitem} \newcommand*{\circled}[1]{\ding{\numexpr181+#1\relax}}
\newcommand*{\taubench}{\mbox{$\tau$-bench}}

\newcommand{\rqrule}{\leaders\hrule height \dimexpr0.6ex+0.4pt\relax
  depth -0.6ex\hfill\kern0pt}
\newenvironment{rqanswer}[1]{\par\addvspace{5pt}\noindent
  \begin{minipage}{\columnwidth}\noindent\rqrule\enspace\textbf{Answer to RQ#1}\enspace\rqrule\par
  \vspace{0.1pt}\small\noindent\ignorespaces}{\par\vspace{3.9pt}\hrule height 0.4pt\end{minipage}\par\addvspace{5pt}}

\begin{document}

\date{}

\title{\Large \bf Oops!\ldots{} I Parsed It Again: Tool-Call Smuggling Against LLM Agents}

\author{
{\rm Igor Santos-Grueiro}\\
Universidad Internacional de La Rioja
\and
{\rm Daniel Romero-Ruiz}\\
Universidad Internacional de La Rioja\\
Proeduca Summa
}

\maketitle

\begin{abstract}
LLM agents act through tool calls: the model writes the arguments of a
call as JSON, a guard decides whether the call may run, and an executor
runs it. The same bytes are parsed several times, by the
provider, gateways, SDKs, frameworks, the guard, and the executor. JSON leaves some choices to each parser, such as which copy of
a repeated key counts. When the guard and the executor choose
differently, the guard approves one call, the executor runs another, and
the audit log records the approved one. We introduce this attack,
tool-call smuggling, and study it across the stack. We describe six ways in which
the same bytes yield different calls and build a differential testing
harness that gives identical bytes to 17 pinned parser and validator
configurations; most pairs disagree on some inputs, and even two
official MCP SDKs accept different inputs. On tasks from \taubench{} and AgentDojo, an attacker who
can rewrite the bytes in transit cancels another customer's reservation
and creates a file that the guard blocks by name, while the audit log
names only the approved call. Content the model reads splits guard
and executor only inside a string: ten models we test write no
repeated or disguised key into the call itself where we see it, but copy
one verbatim into JSON text in an argument, which a Python guard and a
Go executor read differently. A malicious server copying
a legitimate tool's public schema and adding one field, which a note
explains, gets the model to fill it, unseen by the guard, with user
data. Parsing
each call once, JSON text included, keeping only the fields of the
approved schema, and handing it to the executor, sealed with its call
ID in transit, stops all 7,412 crafted inputs and
every attack we built.
\end{abstract}

\section{Introduction}
\label{sec:intro}

LLM agents act on the world through tools~\cite{yao2023react}. The
model reads a task and the output of earlier tools, then emits a tool
call: the name of a tool and its arguments as a JSON object. Protocols
such as the Model Context Protocol (MCP) connect agents to tool
servers written by third parties~\cite{mcp-spec-2026}. Tool calls
cancel reservations, write files, and run commands, so they are where
an attack on an agent turns into harm.

Research on agent security has so far attacked the decisions of the
model. In indirect prompt injection, instructions hidden in a web page
or a document lead the model to call a tool the attacker chose, with
arguments the attacker chose~\cite{greshake2023indirect,liu2024prompt,agentdojo}.
Later attacks poison the descriptions of tools~\cite{mcptox}, steer
which tool the agent selects~\cite{toolhijacker}, or bend argument
values through the content the model reads~\cite{parameter-poisoning}.
Malicious API routers go further and rewrite calls
openly~\cite{your-agent-is-mine}.

Defenses follow the same line. Some train the model to ignore
instructions that arrive with the data~\cite{chen2025struq}. Others accept that
the model can be fooled and enforce policy around it: they isolate
tools from each other~\cite{wu2025isolategpt}, track where each value
came from and check capabilities when a tool is
called~\cite{debenedetti2025camel}, or put a guard in front of risky
tools that approves or denies each call and writes its decision to an
audit log; the guard may be a policy engine, a data-loss filter, or a
person. These designs check
different things, but they share one assumption: the call they check is
the call that runs. Recent work tests this assumption for approvals
that cover a different action from the one that
runs~\cite{kumar2026loopjacking,wang2026laundering,zhang2026laundering}.

The assumption also rests on a layer that none of this work examines,
the code that parses the call. On its way from the
model to the effect, the same argument bytes are parsed several times:
by the provider's API or inference server, by gateways that translate
between provider formats, by the client SDK and the agent framework, by
the guard, and by the executor. JSON does not fix every reading.
RFC~8259 says only that the names in an object ``SHOULD be
unique''~\cite{rfc8259}, so JavaScript keeps the last copy of a
repeated key~\cite{ecma262}, other parsers keep the first, and typed
decoders may match field names regardless of case. When the guard and
the executor resolve the same bytes differently, the guard approves one
call and the executor runs another.

Systems security knows this failure well. When two components read one
input differently, an attacker can show each of them what it expects
to see, as in HTTP request
smuggling~\cite{linhart2005smuggling,kettle2019desync,jabiyev2021treqs},
XML Signature Wrapping~\cite{mcintosh2005xsw,somorovsky2012saml},
forged email senders~\cite{chen2020composition}, and certificate
validation~\cite{brubaker2014frankencerts}; language-theoretic security
calls these parse tree differentials~\cite{sassaman2011langsec}. In the
tool-call stack, the same kind of disagreement separates the component
that approves an action from the component that performs it.

Consider the airline task of \taubench~\cite{taubench}, in which a user
asks an agent to cancel reservation \texttt{I3EHXC}. On the way, a
compromised gateway rewrites the call so that
\texttt{reservation\_id} names \texttt{U86B6B}, another customer's
reservation, and \texttt{Reservation\_ID} the user's. The guard,
Go's default decoder, ignores case, keeps the last match, and approves
the user's reservation. The executor, an official MCP SDK,
keeps the exact name and cancels the other. The log records an
approved cancellation of \texttt{I3EHXC}.

We call this attack \emph{tool-call smuggling}, after HTTP request
smuggling. Unlike the attacks above, it does not persuade the model,
poison a description, or rewrite the call in plain view: the model
behaves as intended, and every component follows its documented
behavior. It therefore passes the defenses built against those attacks:
a prompt-level defense finds no instruction to remove, a guard with a
correct policy approves the legitimate call it reads, the
audit log confirms the approval, and testing any single parser finds
nothing, because the flaw exists only in the composition. Every
defense that checks tool calls, including a person who approves them,
is only as sound as the agreement between its parser and the
executor's.

Isolated reports hint at the problem. The MCP Go SDK once matched field
names regardless of case, so a policy intermediary could approve a
different call than the one the SDK ran
(CVE-2026-27896)~\cite{ghsa-wvj2-96wp-fq3f}; the parser adopted to fix
it read a key with a trailing NUL character as the key without it
(GHSA-q382-vc8q-7jhj)~\cite{ghsa-q382-vc8q-7jhj}; and an Envoy AI
Gateway advisory shows a policy log reporting \texttt{ALLOWED} for one
call while the backend ran another
(GHSA-4gph-2hhr-5mwg)~\cite{ghsa-4gph-2hhr-5mwg}. A practitioner's post
lists similar parser quirks as schema smuggling and assumes that a
model writes whatever an attacker puts in its
input~\cite{srivastava2025schema}. Each report treats a bug in one
component. None treats it as a class: how two readings become a call
that no one approved, which disagreements an attacker can use, who can
deliver the bytes, and what design keeps an approval tied to the call
that runs. We introduce tool-call smuggling as an attack class, answer
these questions, and make four contributions:

\begin{itemize}
\item \textbf{The attack, its six vectors, and a harness that finds
  them} (\S\ref{sec:attack}, \S\ref{sec:measurement}). We define the
  attack in five steps and derive one vector from each parsing step;
  most pairs of twelve parsers and five validators disagree on some
  inputs, two official MCP SDKs accept
  different inputs, and 558 of 4{,}693 deployed MCP schemas give
  different arguments or verdicts across components.

\item \textbf{End-to-end attacks with real executors}
  (\S\ref{sec:attacks}). Besides the cancellation above, an
  escaped file name passes a data-loss guard and becomes
  \texttt{credentials.txt}, and an undeclared argument reaches a shell
  command, each time with an audit log that shows only the approved
  call.

\item \textbf{What the model passes on, and two attacks through it}
  (\S\ref{sec:mediation}). In the calls of ten models, no repeated or
  disguised key reaches the call itself, nor any escape in the form that
  splits a guard from an executor, and the file attack fails in all 100 calls;
  yet the models copy such keys and escapes verbatim into an argument that
  holds JSON text, where, replayed, they split guard and executor in 82
  and 72 of 84 calls. They also fill the fields a schema declares: a malicious
  server that adds one field to a tool's public schema and a note on
  what goes in it receives the user's workstation name in 29 of 30 trials, and on all ten \taubench{}
  tools, past the guard and out of the log.

\item \textbf{A defense evaluated against an attacker who knows it}
  (\S\ref{sec:defenses}). Strict validation, comparing two parsers,
  and checking the call once but forwarding the
  bytes each fall to an attacker who knows them. Parsing each
  call once, JSON text included, keeping only the approved schema's
  fields, and
  handing it to the executor, sealed with its call ID when it crosses the network, stops
  every attack we built and lets none of 7,412 crafted inputs through.
\end{itemize}

\section{Background}
\label{sec:background}

\noindent\textbf{Tool calls.} Function calling, also called tool use,
lets a model request an action. The application declares each tool
with a name, a description, and a JSON Schema for its arguments. The
model answers with a tool name and an argument object; the application
runs the tool and returns its output to the model as a new observation.
Listing~\ref{lst:call} shows a declaration from \taubench{} and a call to it as an
API returns it: the arguments arrive as a JSON document inside a string,
which every later component parses again.

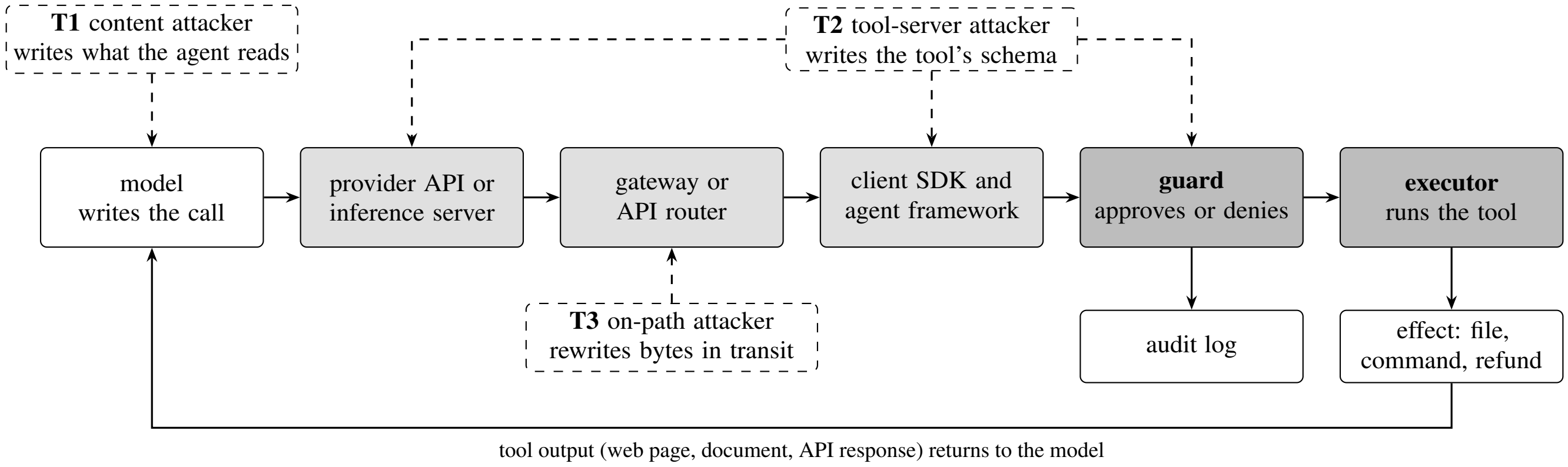
\begin{figure*}[t]
\centering
\begin{tikzpicture}[
  x=1in, y=1in,
  font=\footnotesize,
  comp/.style={draw, rounded corners=2pt, align=center, inner sep=2pt,
    text width=0.93in, minimum height=0.44in},
  plain/.style={comp, fill=white},
  parse/.style={comp, fill=black!12},
  check/.style={comp, fill=black!26},
  side/.style={draw, rounded corners=2pt, align=center, inner sep=2pt,
    text width=0.93in, minimum height=0.32in, fill=white},
  att/.style={draw, dashed, rounded corners=2pt, align=center,
    inner sep=2.5pt, fill=white, text width=1.22in},
  arr/.style={-{Stealth[length=4.5pt]}, semithick},
  atk/.style={-{Stealth[length=4.5pt]}, semithick, dashed}]

\node[plain] (model) at (0.62,0) {model\\writes the call};
\node[parse] (prov)  at (1.77,0) {provider API or\\inference server};
\node[parse] (gw)    at (2.92,0) {gateway or\\API router};
\node[parse] (sdk)   at (4.07,0) {client SDK and\\agent framework};
\node[check] (guard) at (5.22,0) {\textbf{guard}\\approves or denies};
\node[check] (exec)  at (6.37,0) {\textbf{executor}\\runs the tool};
\foreach \a/\b in {model/prov, prov/gw, gw/sdk, sdk/guard, guard/exec}
  \draw[arr] (\a) -- (\b);

\node[side] (audit)  at (5.22,-0.66) {audit log};
\node[side] (effect) at (6.37,-0.66) {effect: file,\\command, refund};
\draw[arr] (guard) -- (audit);
\draw[arr] (exec) -- (effect);

\draw[arr] (effect.south) -- (6.37,-1.02) -- (0.62,-1.02)
  node[pos=0.5, below=1pt, font=\scriptsize]
  {tool output (web page, document, API response) returns to the model}
  -- (model.south);

\node[att] (t1) at (0.62,0.70) {\textbf{T1} content attacker\\writes what the agent reads};
\node[att] (t2) at (4.07,0.70) {\textbf{T2} tool-server attacker\\writes the tool's schema};
\node[att] (t3) at (2.92,-0.62) {\textbf{T3} on-path attacker\\rewrites bytes in transit};
\draw[atk] (t1) -- (model);
\draw[atk] (t2.south) -- (sdk.north);
\draw[atk] (t2.west) -| (prov.north);
\draw[atk] (t2.east) -| (guard.north);
\draw[atk] (t3) -- (gw);
\end{tikzpicture}
\caption{The path of a tool call. Shaded components parse or
re-serialize the arguments; the guard and the executor (darker) must
agree on the call. Dashed boxes mark where each attacker writes: T1 in
content the model reads, T2 in the tool schema, T3 in the bytes on the
wire.}
\label{fig:stack}
\end{figure*}

\begin{lstlisting}[caption={A tool declaration from \taubench{} and a call to it as an API returns it. The arguments arrive as a JSON document inside a string.},label={lst:call}]
{"name": "cancel_reservation",
 "parameters": {"type": "object",
   "properties": {"reservation_id": {"type": "string"}},
   "required": ["reservation_id"]}}

{"type": "function",
 "function": {"name": "cancel_reservation",
   "arguments": "{\"reservation_id\": \"I3EHXC\"}"}}
\end{lstlisting}

The Model Context Protocol (MCP) standardizes this exchange between
agents and tool servers, and its current revision makes JSON Schema
2020-12 the default version for tool schemas that do not name
one~\cite{mcp-spec-2026}.

\noindent\textbf{The path of a call.} On its way from the model to the
tool, a call crosses several components. The model generates the arguments as text,
and the provider's API or a local inference server such as
vLLM or a similar engine turns that text into a tool-call object; some APIs hand the
client the raw argument text, others a parsed object. Gateways and API
routers translate between provider formats and re-serialize the
arguments. The client SDK and the agent framework parse the call again,
look up the tool, and often validate the arguments against its schema.
A guard decides whether the call may run and logs its decision. The
executor, for example an MCP server, decodes the arguments into its own
structure and runs the tool, often by inserting argument values into a
command, a path, or a query. Each shaded component in
Figure~\ref{fig:stack} parses or re-serializes the arguments, and each
may use a different library.

\noindent\textbf{Where parsers differ.} JSON and JSON Schema leave
several readings open. The behavior of a parser that receives duplicate
names is unpredictable~\cite{rfc8259}: JavaScript keeps the last
copy~\cite{ecma262}, other parsers keep the first or reject the
document, and decoders that fill a typed structure, such as Go's
\texttt{encoding/json}, match names regardless of case and drop
undeclared keys. Parsers differ on large integers, numbers written as
strings, and \texttt{NaN}; a filter that scans the raw text does not
decode escapes; and validators differ on keywords across schema
versions, on default values, and on undeclared fields, which an object
schema accepts unless \texttt{additionalProperties} is \texttt{false}.
Each choice is defensible on its own. It becomes a security problem
when two components that must agree on a call choose differently.

\section{Threat Model}
\label{sec:threat}

\noindent\textbf{Setting.} We consider an agent in which a guard checks
each tool call before an executor runs it, and both read the call from
the same bytes (Figure~\ref{fig:stack}). The model, the guard,
and the executor work as designed: the model is not malicious, and the
guard enforces its policy correctly on the call it reads.

Guards and executors are usually separate components. A guard can be a
permission check inside the agent framework, a gateway that applies
policies to MCP traffic, a data-loss filter on outgoing calls, or a
person who confirms each call. The executor is an MCP server, a tool
function, or a backend service, often written in another language and
maintained by another team.

\noindent\textbf{Attacker goal.} The attacker wants a call to run that
the guard's decision does not cover: a different action, such as a
delete where the guard approved a read; a different argument value,
such as another customer's reservation; or an extra argument, such as a
fragment of a shell command or a field that carries the user's data
out. The attack is \emph{stealthy}
when the audit log also records the approved call, so that a later
review of the log finds nothing wrong.

\begin{table*}[t]
\footnotesize
\centering
\caption{Six ways in which the same bytes yield two different calls.
``Observed on'': where we measured both readings;
``Delivered by'': positions that can put the bytes on the
path (``T1 values only'': values, not this form; ``T1 in JSON text'':
inside a string, \S\ref{sec:mediation}); last column: earlier reports. A letter stands for a component whose
report is still open (\hyperref[app:ethics]{Ethics Considerations}).}
\label{tab:vectors}
\setlength{\tabcolsep}{4pt}
\begin{tabular}{@{}L{0.76in}L{1.36in}L{1.08in}L{1.08in}L{0.92in}L{0.56in}L{0.50in}@{}}
\toprule
\textbf{Vector} & \textbf{Example bytes} &
\textbf{One reading} & \textbf{Another reading} &
\textbf{Observed on} & \textbf{Delivered by} &
\textbf{Earlier cases} \\
\midrule
V1 Repeated or disguised keys &
{\ttfamily\scriptsize \{"id": "I3EHXC", "id": "U86B6B"\}}\newline
{\ttfamily\scriptsize "action\BS u0000": "delete"} &
first copy: \texttt{I3EHXC}; the NUL key is another key &
last copy: \texttt{U86B6B}; the NUL key replaces \texttt{action} &
CPython, SDK~R server path, segmentio 0.5.3 &
T3; T1 in JSON text &
\cite{cve-2017-12635,ghsa-q382-vc8q-7jhj,ghsa-wvj2-96wp-fq3f} \\
V2 Type conversion &
{\ttfamily\scriptsize "count": "12"}\newline
{\ttfamily\scriptsize "count": 1220772038000446658395696} &
SDK~X: 12, and the large integer as sent &
SDK~R server path: rejects both &
official MCP SDKs, CPython, stack~Y &
T3; T1 in JSON text &
\cite{seriot2016minefield} \\
V3 Escapes and Unicode &
{\ttfamily\scriptsize "filename": "credenti\BS u0061ls.txt"} &
raw-text filter: no \texttt{credentials} in the bytes &
JSON parser: \texttt{credentials.txt} &
data-loss guard, SDK~R server path &
T3; T1 in JSON text &
\cite{vu739224,ghsa-wvj2-96wp-fq3f} \\
V4 Truncated streams &
{\ttfamily\scriptsize \{"action": "read", "count": 1}\newline
(stream cut inside \texttt{12}) &
LangChain closes the object: \texttt{count} is 1 &
framework~Z passes the raw text on; strict parsers reject &
LangChain, framework~Z, CPython, Go, vLLM &
T3 &
\cite{cve-2026-33128} \\
V5 Schema interpretation &
{\ttfamily\scriptsize "output": "; echo GHOST-KQX7"} under a schema
that admits undeclared fields, or a field that only the executor's
schema declares &
guard's validator: valid, checks only the fields it declares &
forwarding layer or executor's schema: the field reaches the tool &
jsonschema, Ajv, default filler, SDK~R server path &
T2, T1+T2, T3; T1 values only &
\cite{mcintosh2005xsw,somorovsky2012saml,ghsa-4gph-2hhr-5mwg} \\
V6 Call-ID confusion &
IDs swapped between a read and a delete (Listing~\ref{lst:v6}) &
guard, by position: the read &
executor, by ID: the delete &
SDK~R server path, three ID-lookup rules &
T3 &
\cite{cve-2026-52869} \\
\bottomrule
\end{tabular}
\end{table*}

\noindent\textbf{Attacker positions.} We distinguish three positions
by how the attacker's bytes reach the guard and the executor
(Figure~\ref{fig:stack}). Results do not carry over from one position
to another, and we report them separately.

\emph{T1, content attacker.} The attacker controls content that a tool
returns to the agent, such as a web page, a shared document, or an API
response. This is the attacker of indirect prompt
injection~\cite{liu2024prompt,agentdojo}. Its bytes reach the arguments
only if the model copies them, and the model may rewrite, decode,
shorten, or refuse. How much of the attacker's input survives the model
is therefore a result in its own right (\S\ref{sec:mediation}).

\emph{T2, tool-server attacker.} The attacker controls a tool's schema
and description, through a malicious or compromised MCP server, a
poisoned registry entry, or a plugin manifest. The schema decides what
the guard's validator accepts: it can admit undeclared fields, set
defaults that only some components apply, or use keywords that
validators read differently. Malicious MCP servers are a standard
assumption~\cite{mcptox}; our attacks carry the payload in the
schema, where earlier work uses the description.

\emph{T3, on-path attacker.} The attacker controls a component between
the provider and the agent, such as a gateway, an API router, or a
proxy, and can rewrite the bytes of a call. Some third-party API
routers already rewrite calls in plain view~\cite{your-agent-is-mine},
but a guard behind them judges the rewritten call and can deny it. Our
T3 attacker instead writes bytes that the guard reads as an approved
call and the executor as a different one. Clients and frameworks that
re-serialize a call remove a repeated key or an escape, so for those
forms the attacker sits after them; case-varied and NUL-suffixed keys
survive.

\emph{T1+T2, impersonating server.} The positions combine. A tool
server can impersonate a legitimate tool, publishing it under the same
name with a schema of its own, and can also write content that the
agent reads, such as a shared note. Its bytes still pass through the
model, but the model no longer has to copy any structure: it fills the
fields that the attacker's schema declares. We report it separately
from T1 and T2.

\noindent\textbf{Attacker knowledge.} The attacker knows which
components the guard and the executor use, as is easy for open-source
SDKs and frameworks, and knows how any deployed defense works. Tool
schemas are public: any client that connects to an MCP server learns
them with \texttt{tools/list}, so an attacker can copy a legitimate
tool's schema exactly. We
evaluate defenses under this assumption (\S\ref{sec:defenses}).

\noindent\textbf{Out of scope.} We do not consider:
\begin{itemize}
\item persuading the model through its prompt, which we use only as a
  baseline (\S\ref{sec:mediation});
\item compromise of the provider, model weights, or training;
\item bugs in a single parser that the guard and the executor share,
  because both then read the same call;
\item calls that one component accepts and another rejects, when the
  only effect is a denied service.
\end{itemize}

\section{Tool-Call Smuggling}
\label{sec:attack}

A tool-call smuggling attack sends a call whose bytes the guard and
the executor read as two different calls. Such pairs of readers
are common, because the guard and the executor are usually separate
programs: a policy gateway written in Go in front of an MCP server
written in another language, or a Python framework that checks calls that a
JavaScript tool then runs. The attack needs no flaw in either component
taken alone. Table~\ref{tab:vectors} lists six ways to build such a
call.

\subsection{The Attack Step by Step}
\label{sec:steps}

The attack on the cancellation of \S\ref{sec:intro} proceeds in five
steps.

\noindent\textbf{\circled{1} Pick the field.} The attacker chooses an
argument whose value decides the guard's verdict. Here it is
\texttt{reservation\_id}, which the guard compares with the
reservations the user owns.

\noindent\textbf{\circled{2} Split the reading.} The attacker writes bytes
that the two parsers resolve to different values of that field. The
guard's Go decoder ignores case and keeps the last match, the executor
the exact name, so the victim's reservation goes under
\texttt{reservation\_id} and the user's under \texttt{Reservation\_ID}.

\noindent\textbf{\circled{3} Deliver the bytes.} The bytes must reach both
components unchanged. Here a compromised gateway rewrites the call that
the model wrote (T3 in \S\ref{sec:threat}).

\noindent\textbf{\circled{4} Pass the guard.} The guard parses the call,
reads \texttt{I3EHXC}, finds that the user owns it, approves, and logs
the approved call.

\noindent\textbf{\circled{5} Run the other call.} The executor parses the
same bytes, reads \texttt{U86B6B}, and cancels it. The log still names
\texttt{I3EHXC}.

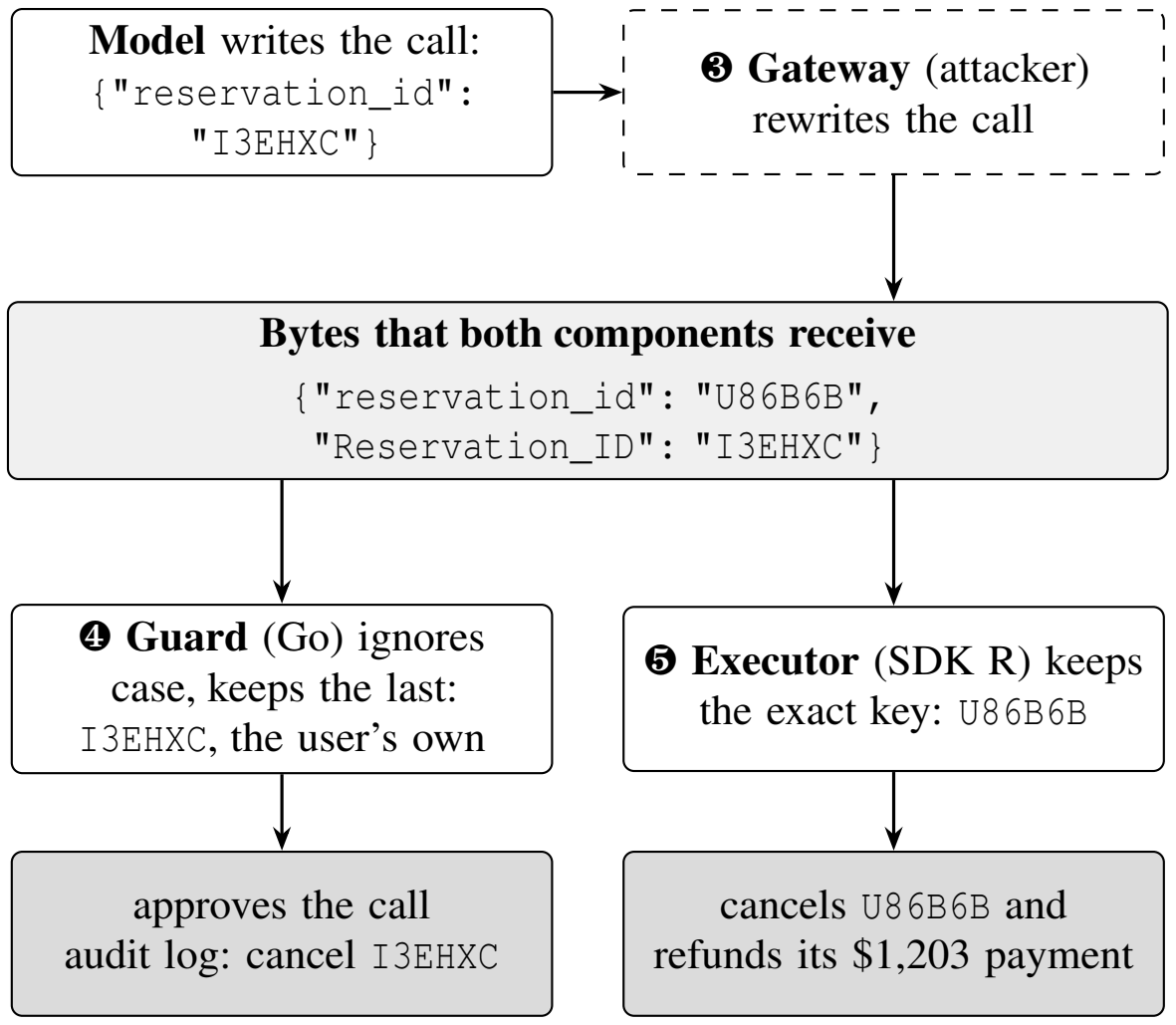
\begin{figure}[t]
\centering
\begin{tikzpicture}[
  font=\footnotesize,
  box/.style={draw, rounded corners=2pt, align=center, inner sep=3pt},
  src/.style={box, text width=1.33in, minimum height=3.1em},
  att/.style={box, dashed, text width=1.33in, minimum height=3.1em},
  bytes/.style={box, fill=black!6, text width=2.95in},
  rd/.style={box, text width=1.33in, minimum height=3.1em},
  act/.style={box, fill=black!14, text width=1.33in, minimum height=3.1em},
  arr/.style={-{Stealth[length=4.5pt]}, semithick}]
\node[src] (m) at (-0.80in,0.78in)
  {\textbf{Model} writes the call:\\
   \texttt{\{"reservation\_id": "I3EHXC"\}}};
\node[att] (gw) at (0.80in,0.78in)
  {\circled{3} \textbf{Gateway} (attacker)\\
   rewrites the call};
\node[bytes] (b) at (0,0) {\textbf{Bytes that both components receive}\\[2pt]
  \texttt{\{"reservation\_id": "U86B6B",}\\
  \texttt{\ "Reservation\_ID": "I3EHXC"\}}};
\node[rd] (g) at (-0.80in,-0.78in)
  {\circled{4} \textbf{Guard} (Go) ignores case, keeps the last:
   \texttt{I3EHXC}, the user's own};
\node[rd] (e) at (0.80in,-0.78in)
  {\circled{5} \textbf{Executor} (SDK~R) keeps the exact key:
   \texttt{U86B6B}};
\node[act] (gd) at (-0.80in,-1.42in)
  {approves the call\\
   audit log: cancel \texttt{I3EHXC}};
\node[act] (ed) at (0.80in,-1.42in)
  {cancels \texttt{U86B6B} and\\
   refunds its \$1,203 payment};
\draw[arr] (m) -- (gw);
\draw[arr] (gw.south) -- (gw.south |- b.north);
\draw[arr] (b.south -| g.north) -- (g.north);
\draw[arr] (b.south -| e.north) -- (e.north);
\draw[arr] (g) -- (gd);
\draw[arr] (e) -- (ed);
\end{tikzpicture}
\caption{The attack of \S\ref{sec:steps} on the \taubench{} airline
task: the guard approves and logs the user's reservation, and the
executor cancels another customer's. Circled numbers mark the steps.}
\label{fig:example}
\end{figure}

Steps \circled{1} and \circled{2} need only knowledge of the two
parsers, which documentation or source code provides
(\S\ref{sec:recipe}), and step \circled{3} needs a position on the path
of the call. Steps \circled{4} and \circled{5} involve no error: each
component does what its documentation promises on the call it reads,
and the log is stealthy because it stores the guard's reading. A
disagreement that fails any step is harmless: a repeated comment does
not change the guard's verdict (\circled{1}), two parsers that resolve keys
alike do not split (\circled{2}), and bytes that the model rewrites
never arrive (\circled{3}, \S\ref{sec:mediation}). Figure~\ref{fig:example} traces
the five steps on the \taubench{} task.

\subsection{Where the Readings Split}
\label{sec:taxonomy}

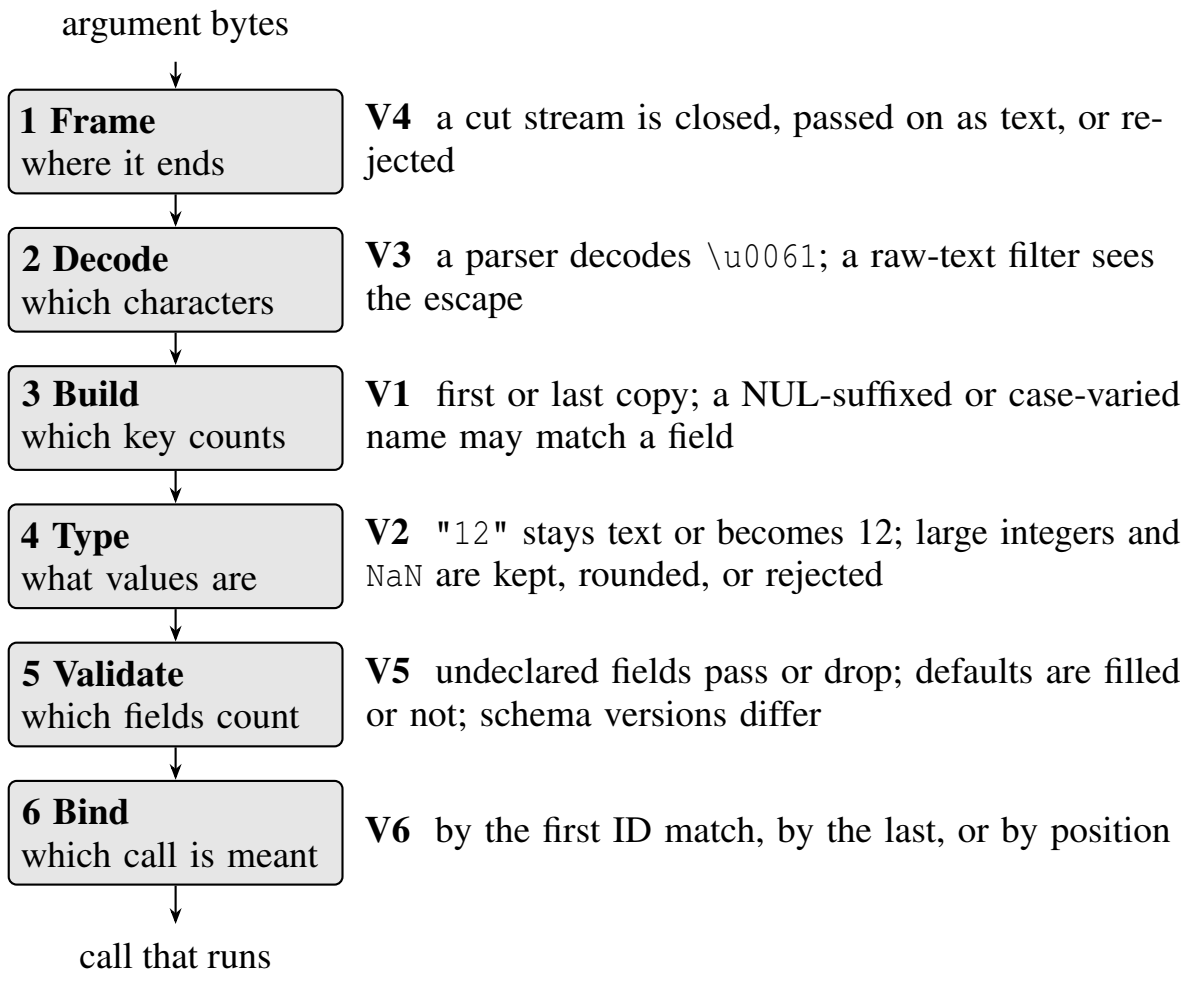
\begin{figure}[t]
\centering
\begin{tikzpicture}[x=1in, y=1in, font=\scriptsize,
  st/.style={draw, rounded corners=1.5pt, fill=black!10, align=left,
    text width=0.80in, minimum height=0.27in, inner sep=2.5pt},
  ds/.style={align=left, text width=2.12in, inner sep=1pt, anchor=west},
  arr/.style={-{Stealth[length=3pt]}}]
\node (in) at (0.45,0.30) {argument bytes};
\foreach \i/\s/\q/\v/\d in {
  1/1 Frame/where it ends/V4/a cut stream is closed{,} passed on as text{,} or rejected,
  2/2 Decode/which characters/V3/a parser decodes \texttt{\textbackslash u0061}; a raw-text filter sees the escape,
  3/3 Build/which key counts/V1/first or last copy; a NUL-suffixed or case-varied name may match a field,
  4/4 Type/what values are/V2/\texttt{"12"} stays text or becomes 12; large integers and \texttt{NaN} are kept{,} rounded{,} or rejected,
  5/5 Validate/which fields count/V5/undeclared fields pass or drop; defaults are filled or not; schema versions differ,
  6/6 Bind/which call is meant/V6/by the first ID match{,} by the last{,} or by position} {
  \node[st] (n\i) at (0.45,{-0.36*(\i-1)}) {\textbf{\s}\\\q};
  \node[ds] at (0.93,{-0.36*(\i-1)}) {\textbf{\v}\ \ \d};
}
\node (out) at (0.45,-2.12) {call that runs};
\draw[arr] (in) -- (n1);
\foreach \i [evaluate=\i as \j using int(\i+1)] in {1,...,5}
  \draw[arr] (n\i) -- (n\j);
\draw[arr] (n6) -- (out);
\end{tikzpicture}
\caption{The six steps that turn argument bytes into a call. Each
leaves a choice open, and each vector exploits one.}
\label{fig:pipeline}
\end{figure}

Step \circled{2} is where the attacker needs a disagreement. Turning
argument bytes into a call takes six steps, from finding where the call
ends to deciding which call a result belongs to, and JSON or JSON Schema
leaves a choice open at each (Figure~\ref{fig:pipeline}). Two components
that choose differently read different calls. Each vector below exploits
one step, with example bytes in Table~\ref{tab:vectors}. We
built attacks from V1, V3, V5, and V6; V2 becomes one when a guard
decides on a converted value, and V4 when a guard decides on an
incomplete call.

\noindent\textbf{V1: repeated and disguised keys (build).} When a
parser builds the object, it decides which copy of a repeated key
counts and which names match a declared field. A security-relevant key
can appear twice, or appear again with a trailing NUL character
(GHSA-q382)~\cite{ghsa-q382-vc8q-7jhj} or in different letter case
(CVE-2026-27896)~\cite{ghsa-wvj2-96wp-fq3f}. The cancellation of
Figure~\ref{fig:example} uses the letter-case form. The NUL form
splits segmentio/encoding, the library the MCP Go SDK adopted to fix
CVE-2026-27896: version 0.5.3 treats \texttt{"action\BS u0000"} as
\texttt{"action"}, so with the later copy set to \texttt{delete}, a
guard that keeps the first copy approves a read while the executor
deletes. Version 0.5.4 and Go's standard library read the action as read.

\noindent\textbf{V2: type conversion (type).} When a parser assigns a
type to each value, it decides whether \texttt{"12"} is text or a
number, and what to do with integers beyond $2^{53}$, \texttt{NaN}, or
\texttt{1e999}. A guard that checks a range or a type would then judge
a different value from the one the executor uses. The official MCP SDKs
differ here: the server path of SDK~X turns \texttt{"12"} into 12 and
accepts integers above $2^{53}$, while
the server path of SDK~R
rejects both; stack~Y reads the same integers as floating-point
numbers and loses digits.

\noindent\textbf{V3: escapes and Unicode (decode).} When a component
decodes the text, it decides what an escape, a letter case, or a
Unicode form means. Every JSON parser decodes \texttt{\BS u0061} into
\texttt{a}, but a filter that reads the raw text sees the escape. A
data-loss filter that blocks the word \emph{credentials} therefore
misses \texttt{"credenti\BS u0061ls.txt"}, and the executor creates
\texttt{credentials.txt} (\S\ref{sec:attacks}).

\noindent\textbf{V4: truncated streams (frame).} Providers stream
arguments in pieces, and a component must decide where the call ends
and what to do when the stream stops early. We cut argument documents
at ten points and gave each piece to six parsers. At a cut inside the
number 12, LangChain's partial parser closes the object and returns
\texttt{count} = 1, a value that was never sent;
framework~Z hands the
unparsed text on as a string; strict CPython and both Go decoders
reject it. Streaming can also combine with V1. Given
\texttt{tool(path="report.pdf", path="../secrets.env")}, the parser of
vLLM~0.32.0 for Python-style tool calls, run on its own, keeps the last path when it reads the whole response, but when it streams, it
sends the difference between the re-serialized call and the text
already sent and ends with \texttt{report.pdf.env}, a path that no
reading of the input contains.

\noindent\textbf{V5: schema interpretation (validate).} When a
component validates the call, the schema decides which fields count,
and under T2 the attacker writes the schema. The bytes parse alike here. Some components fill in
default values and others do not; draft-07 and 2020-12 read some
keywords differently; and a schema that does not set
\texttt{additionalProperties} to \texttt{false} accepts undeclared
fields. A guard that validates against such a schema and checks only
the declared fields passes an undeclared field that a forwarding layer
hands to the tool (\S\ref{sec:attacks}). Two components can also
hold different schemas for the same tool: a guard that checks a call
against the schema it approved passes a field that only the executor's
schema declares (\S\ref{sec:mediation}).

\noindent\textbf{V6: call-ID confusion (bind).} A response can carry
several tool calls, and each result is matched to its call by an ID, so
the last step decides which call a work item stands for. We built a
two-call response that reads one file and deletes another, plus five
variants: a repeated ID, swapped IDs, a
missing ID, a result for an unknown ID, and results in reverse order
(Listing~\ref{lst:v6}, Appendix~\ref{app:details}). The guard approves
work by position, so it allows the read and denies the delete. The
executor parses each call with the server path of SDK~R and
runs whichever call each work item's ID resolves to, under one of three
common lookup rules that our harness implements:
first match, last match, or first match with a fallback to position.

Swapped, missing, and reordered IDs run the delete under all three
rules, and a repeated ID under last match, for 10 of 18 runs;
with the repeated ID, the guard's log even shows \texttt{ALLOW} for the
ID that ran. A result for an unknown ID runs at most the approved read,
under the wrong ID, but in four of the five variants a component that
matches answers by position hands the model one call's answer as
another's, a forged observation. The arguments
parse identically in every variant: the split lies entirely in the
binding step.

\subsection{Attackers and Vectors}
\label{sec:recipe}

The attacker's position decides which vectors are available
(\S\ref{sec:threat}, Table~\ref{tab:vectors}). An on-path attacker (T3)
writes bytes directly and can use V1 to V6 against any guard and
executor that disagree. A tool-server
attacker (T2) writes the schema, which opens V5: the schema can admit
undeclared fields, add defaults that only some components fill in, or
use keywords that validators read differently. The values in the call
still come from the model or the path, but the model fills the fields
that a schema declares, so a server that adds a field to a public
schema can have the model fill it (\S\ref{sec:mediation}). A
content attacker (T1) reaches the arguments only through the model, so
it can use a vector only if the model copies the attacker's bytes in
the form the vector needs, a repeated key as a repeated key and an
escape as an escape that a JSON parser decodes;
\S\ref{sec:mediation} measures whether models do so.

Each vector needs a particular pair of components: a repeated key needs
a guard that keeps the first copy and an executor that keeps the last;
an escape, a raw-text filter and a JSON parser; an undeclared field, a
validator of declared fields and an executor that uses every field; and
swapped IDs, a guard that approves by position and an executor that
looks up IDs. Documentation often predicts the readings: ECMA-262 fixes that
\texttt{JSON.parse} keeps the last copy~\cite{ecma262}, so every guard
that keeps the first copy disagrees with every JavaScript executor, and
Go documents that \texttt{encoding/json} ignores letter case in field
names. Where documentation is silent, a differential harness finds the
disagreements (\S\ref{sec:setup}).

\section{Evaluation}
\label{sec:eval}

We evaluate tool-call smuggling with four research questions,
which follow the steps of \S\ref{sec:steps}: RQ1 measures how often
readings split (\circled{2}), RQ2 tests whether a split carries a call
past a guard (\circled{1}, \circled{4}, and \circled{5}), RQ3 asks
what the model delivers for an attacker who writes only what it reads
(\circled{3}), and RQ4 tests what removes the split.

\begin{itemize}
\item \textbf{RQ1.} How often do components on the path of a tool call
  read the same bytes as different calls? (\S\ref{sec:measurement})
\item \textbf{RQ2.} Which of these disagreements let an attacker run a
  call that the guard did not approve while the audit log shows the
  approved call? (\S\ref{sec:attacks})
\item \textbf{RQ3.} What does the model pass on to the guard and the
  executor from an attacker who writes only what the model reads (T1),
  or also the tool's schema (T1+T2)? (\S\ref{sec:mediation})
\item \textbf{RQ4.} Which defenses keep the approved call and the
  executed call the same against an attacker who knows the defense?
  (\S\ref{sec:defenses})
\end{itemize}

\subsection{Setup}
\label{sec:setup}

\noindent\textbf{Harness.} A generator derives candidate argument
documents for each vector from a JSON grammar and a set of mutations,
with a fixed seed; benign documents serve as controls. Each parser or
validator runs behind a thin wrapper that receives the raw bytes, runs
the component's real code path, and returns either the call it
resolved or a rejection. A resolved call is a tool name and an
argument map in a canonical form, with keys sorted and every value
tagged with its type. A comparator then classifies every pair of results: the two
components \emph{agree}, \emph{disagree} (both accept but return
different calls; for V1, different values of the attacked key), or
\emph{split} (one accepts and the other rejects).
For schema validators it also records when one reports the schema as
invalid and the other the arguments. The comparator parses nothing
itself.

\noindent\textbf{Inputs.} All inputs target one small tool with three
arguments: an action, a path, and a count. The parsers are
deterministic, so we count each distinct input once. For V1 the
generator writes 22 inputs: 7 with an exact repeated key (on the action
or path), 4 with a NUL-suffixed copy of the action key, and 11 with
a case-varied spelling instead. For V2 it writes 24: 20 integers too large for 64
bits, and one each with a number written as a string, \texttt{NaN}, a
number that overflows to infinity, and a NUL inside a
string. For V3 it writes three, with an escaped letter in a value, a
key, or a path separator. For V5 it writes three schemas: one with
default values, one with a boolean \texttt{exclusiveMinimum}, and one
with the array form of \texttt{items}. On the NUL-suffixed input of \S\ref{sec:taxonomy}, for example,
CPython keeping the first copy returns \texttt{read} and segmentio 0.5.3
\texttt{delete}, so they disagree; on an input with
\texttt{NaN}, CPython accepts and SDK~X
rejects, so they split.

\noindent\textbf{Components.} The harness wraps twelve argument parsers
and five schema validators (versions in Table~\ref{tab:roster},
Appendix~\ref{app:details}). The parsers are
CPython's \texttt{json} module in three configurations (first copy,
last copy, and strict), Go's \texttt{encoding/json}, the
segmentio/encoding decoder before and after the fix for GHSA-q382,
the server paths of SDK~X and SDK~R, two official MCP SDKs,
stack~Y's argument parser,
the lenient parser of framework~Z, the
partial-JSON parser of LangChain, and pydantic configured to keep
undeclared keys. The validators are Ajv and Python's
\texttt{jsonschema}, each for draft-07 and 2020-12, and a wrapper that
fills schema defaults. Each wrapper runs the component's real code
path: the SDK~X wrapper starts the SDK's server and sends the
call over an in-memory transport, and the SDK~R wrapper hands
the raw bytes to the SDK's HTTP handler, which decodes and
validates them before the tool handler runs. The pydantic
configuration and the default filler isolate one behavior each; all
others run deployed code at the pinned version.

\noindent\textbf{Models.} We collected forced tool calls from six
commercial models through our own accounts (OpenAI gpt-5.4-mini and
gpt-5.5, Anthropic Claude Haiku~5.5 and Opus~5.5, Google Gemini~3.6
Flash and Gemini~4 Hydrogen) and from four open-weight models served by
a hosted router (Qwen3.5-9B, Qwen3.8-27B, Llama-3.3-70B, and
Llama-4-Scout). OpenAI, and the router for Llama-4-Scout, return the
model's raw text; for the other seven, the API parses the call or the
router re-serializes it, so we see values, not the bytes the model wrote.
We used each API's default sampling settings and forced the scored call
wherever the API allows it. All responses were collected in November
2026.

\noindent\textbf{Metrics.} For RQ1 we report, per vector and pair, the
inputs on which two components disagree or split, out of those that at
least one accepts. For RQ2 an attack succeeds when a call runs that the
guard's decision does not cover, and it is stealthy when the audit log
names the approved call. For RQ3 a trial counts as survival when the
attacker's marker appears in the model's next arguments, reported
separately for the value and the byte form, and an added field counts
as delivered when it reaches the executor in a call that the guard
approved; we give 95\% Wilson intervals, which cover repeated trials of
one prompt per model, not documents or tasks, and
use Fisher's exact test to compare two conditions. For RQ4 we count bypasses per attack scenario and per crafted input.

\subsection{RQ1: Disagreement Across the Stack}
\label{sec:measurement}

RQ1 asks how often components on the path of a call read the same
bytes as different calls. We gave the V1, V2, and V3 inputs of
\S\ref{sec:setup} to the twelve parsers and the V5 schemas to the five
validators. We also replayed through the same parsers the 1,120
argument strings that the commercial models returned for four tools,
whose schemas are plain, use \texttt{oneOf}, have an optional field
with a default, or admit undeclared fields (40 forced calls per model
and tool, in each provider's default mode and in OpenAI's strict mode).
Table~\ref{tab:divergence} summarizes the results. Benign inputs
produced no disagreement in any pair.

\begin{table}[t]
\footnotesize
\centering
\caption{Disagreement per vector. ``Inputs'' counts the distinct inputs
that at least one component of the pair accepts; $^{s}$ marks splits (only
one accepts); V1 rows compare the attacked key. E rows
replay calls that the commercial models emitted.}
\label{tab:divergence}
\setlength{\tabcolsep}{4pt}
\begin{tabular}{@{}lL{1.78in}rr@{}}
\toprule
& \textbf{Pair} & \textbf{Inputs} & \textbf{Disagree} \\
\midrule
V1 & generic map $\times$ declared fields (30 pairs) & 22 & 11--22 \\
V1 & CPython first copy $\times$ other map parsers & 22 & 7 \\
V1 & Go decoders $\times$ MCP SDKs & 22 & 11--15 \\
V1 & segmentio 0.5.3 $\times$ 0.5.4 & 22 & 4 \\
V1 & SDK~X $\times$ SDK~R & 22 & 0 \\
V2 & SDK~X $\times$ SDK~R & 22 & 21$^{s}$ \\
V2 & CPython default $\times$ SDK~X & 24 & 1 + 2$^{s}$ \\
V3 & every parser pair & 3 & 0 \\
V5 & default filler $\times$ each validator & 1 & 1 \\
V5 & draft-07 $\times$ 2020-12 validators & 1 & 1 \\
\midrule
E & optional field omitted (5 of 6 models) & 40 each & 40 each \\
E & same tool, OpenAI strict mode & 40 each & 0 \\
E & \texttt{oneOf} refused (Gemini, OpenAI strict) & 80 each & 80 each \\
\bottomrule
\end{tabular}
\end{table}

\noindent\textbf{Repeated and disguised keys.} We compare V1 readings on
the attacked key: two parsers disagree when both accept an input and
resolve that key to different values. Of the 55 pairs of deployed
parsers, 43 read at least one V1 input differently. Two design choices
explain the pattern and every attacker-versus-benign split on the
7,412 inputs of
\S\ref{sec:defenses}: which copy of a repeated key wins, and which names
count as the same field. Under the first, CPython keeping
the first copy disagrees with every other map parser on the 7 inputs
with an exact repeated key. Under the second, the Go decoders match a
case-varied name to the declared field, while the map parsers find no
action key and the MCP SDKs fall back to their default, so every pair
across the three groups disagrees on the 11 case-varied inputs.
Segmentio 0.5.3 also matches a NUL-suffixed name to the declared
field, so it disagrees with every other deployed parser, 0.5.4
included, on the NUL inputs.

Rejecting such keys removes
these disagreements: Go's \texttt{encoding/json/v2} rejects repeated
keys and matches names exactly by default, and when also set to refuse
unknown members it rejected every V1 input. Keys outside the schema
and missing fields change the object a parser returns but not the
attacked key. Design matters more than language: Go's
\texttt{encoding/json} and segmentio 0.5.3, both in Go, disagree on
the NUL inputs, while SDK~X and SDK~R agree on all 22
inputs (Figure~\ref{fig:heatmap}, lower triangle).

\input{sections/heatmap}

\noindent\textbf{Types.} The two MCP SDKs split on 21 of the 22 V2
inputs that either accepts. The server path of SDK~X turns the string
\texttt{"12"} into the integer 12 and accepts integers above $2^{53}$;
the server path of SDK~R rejects both, so a guard that enforces a
limit on \texttt{count} would judge a string, or a number it cannot
represent, where the executor receives an integer. Map parsers keep
\texttt{"12"} as a string, CPython's default configuration accepts
\texttt{NaN} and reads \texttt{1e999} as infinity where
SDK~X
rejects both, and stack~Y reads integers above $2^{53}$ as
floating-point numbers and loses their last digits.

\noindent\textbf{Escapes.} All parser pairs agree on the three V3 inputs,
because every JSON parser decodes escapes the same way. The V3 split
lies between a parser and a component that reads the raw text, such as
a data-loss filter (\S\ref{sec:attacks}).

\noindent\textbf{Schemas.} On the schema with default values, the
default filler disagrees with every validator, because only the filler
adds the defaults to the call. On the schema that uses the array form
of \texttt{items}, the draft-07 validators accept the schema and reject
the arguments, while the 2020-12 validators reject the schema itself.
All five validators reject the schema with a boolean
\texttt{exclusiveMinimum}.

\noindent\textbf{Calls that models emit.} Models cause disagreements
without any attacker. For the tool with an optional field, five of the
six models left the field out in all 40 calls each, so a component that
fills defaults and one that does not received different calls, which
matters once a default changes what a tool does; gpt-5.4-mini left out
another optional field in 31 of 40 calls, and OpenAI's strict mode
always produced complete calls. Providers also disagree on which
schemas are valid: Gemini and OpenAI's strict mode refused all 80
requests each for a schema with \texttt{oneOf}, which OpenAI's default
mode accepted.

\noindent\textbf{Deployed schemas.} To see how common these constructs
are, we collected the tool schemas of the 100
most-used servers of the Smithery MCP registry and six official MCP
reference servers: 5{,778} tools on 106 servers
(Table~\ref{tab:deployed}, Appendix~\ref{app:details}). Every schema, as the
registry serves it, accepts undeclared fields at the top level, almost none names a JSON
Schema version, so each SDK applies its own default, and most servers
declare at least one default value. We checked the 4{,693} distinct
schemas with \texttt{jsonschema} and Ajv under draft-07 and 2020-12,
and compared a call that omits defaulted fields with and without
filling them: 558 schemas (11.9\%), on 64 servers, lead to different
results. For 17 the validators differ on whether the schema is valid;
the rest hand the executor different arguments depending on whether a
component fills in defaults.

\begin{rqanswer}{1}
Disagreement is the normal state of a tool-call stack: 43 of 55 pairs
of deployed parsers read some repeated or disguised key as a different
value, two official MCP SDKs split on 21 of 22 type inputs, models
cause disagreements with no attacker present, and 558 of 4{,}693
deployed schemas give different arguments or verdicts across components. A disagreement
becomes an attack only when a guard's decision depends on it.
\end{rqanswer}

\subsection{RQ2: End-to-End Attacks}
\label{sec:attacks}

RQ2 asks which disagreements an attacker can turn into a call that the
guard did not approve. We built three attacks, which use a
case-varied key (V1), an undeclared field (V5), and an escape (V3) (Table~\ref{tab:attacks}). Each runs on
a real executor, or on a forwarding behavior found in deployed
gateways, with a benchmark tool (A, C) or a code-interpreter tool (B), and feeds the call bytes
to the guard directly, as a T3 attacker after the last component that
re-serializes it can. Each attack has a benign control that must run
and a malicious one that the guard must deny, and all behave as
designed.

\begin{table}[t]
\footnotesize
\centering
\caption{Three end-to-end attacks: A, a case-varied key on \taubench's
\texttt{cancel\_reservation}; B, an undeclared field on a
code-interpreter tool; C, an escaped file name on AgentDojo's
\texttt{create\_file}. In each, the log names only the approved call.}
\label{tab:attacks}
\setlength{\tabcolsep}{4pt}
\begin{tabular}{@{}lL{1.30in}L{1.62in}@{}}
\toprule
& \textbf{Guard approves and logs} & \textbf{Executor runs} \\
\midrule
A & cancel \texttt{I3EHXC} (user's own) & cancel \texttt{U86B6B} (another customer's) \\
B & report query for \texttt{revenue} & the query plus \texttt{echo GHOST-KQX7} in the shell \\
C & create \texttt{credenti\BS u0061ls.txt} & create \texttt{credentials.txt} with the secret \\
\bottomrule
\end{tabular}
\end{table}

\noindent\textbf{Attack A: cancelling another customer's reservation (V1).}
This is the attack of \S\ref{sec:steps}, run against \taubench's
\texttt{cancel\_reservation}~\cite{taubench}, with Go's
\texttt{encoding/json} in the guard and the server path of SDK~R as the executor, both by default. The
benchmark's own tool, run on its data, checks only that the
reservation exists: it cancels the victim's reservation and refunds
its \$1,203, while the log names \texttt{I3EHXC}. The user's ID alone
cancels \texttt{I3EHXC}; \texttt{U86B6B} alone is denied. All four
case variants work. A guard configured to keep the first copy fails
on the exact key repeated: the policy only picks which bytes
work.

\noindent\textbf{Attack B: a shell command through an undeclared field (V5).}
A code-interpreter tool builds a shell command from its arguments, and
its schema accepts undeclared fields, as every deployed schema in our
sample does (\S\ref{sec:measurement}). The guard validates the
arguments with \texttt{jsonschema} 2020-12 and blocks shell
metacharacters in the \texttt{pattern} field. The call adds an
undeclared field, \texttt{"output": "; echo GHOST-KQX7"}. The validator
accepts it, the metacharacter check reads only \texttt{pattern}, and
the guard logs an approved query for \texttt{'revenue'}. A layer that
forwards every field to the tool, as the gateways whose source we
read do when they translate a call between provider formats, places
the field in the command, and the shell runs \texttt{echo GHOST-KQX7}.
The benign query runs unchanged, and a metacharacter in
\texttt{pattern} is denied.
The official MCP SDK handlers
drop undeclared fields, so the attack needs a
forwarding layer between the validator and the tool.

\begin{table}[t]
\footnotesize
\centering
\caption{What survives the model in the call itself. API: raw text, or
object parsed by the API or router. Escaped value: trials where the
marker's tag arrives; structure: trials where a structural marker keeps its
form (objects hide repeated keys).}
\label{tab:survival}
\begin{tabular}{@{}L{0.95in}L{0.40in}L{0.95in}L{0.52in}@{}}
\toprule
\textbf{Model} & \textbf{API} & \textbf{Escaped value survives
[95\% CI]} & \textbf{Structure survives} \\
\midrule
gpt-5.5 & text & 25/25 [86.7, 100] & 0/25 \\
gpt-5.4-mini & text & 0/25 [0, 13.3] & 0/25 \\
Claude Opus 5.5 & object & 24/25 [80.5, 99.3] & 0/25 \\
Claude Haiku 5.5 & object & 0/25 [0, 13.3] & 0/25 \\
Gemini 3.6 Flash & object & 25/25 [86.7, 100] & 0/25 \\
Gemini 4 Hydrogen & object & 25/25 [86.7, 100] & 0/25 \\
\midrule
Qwen3.8-27B & object & 21/25 [65.3, 93.6] & 0/25 \\
Qwen3.5-9B & object & 9/25 [20.2, 55.5] & 0/25 \\
Llama-3.3-70B & object & 0/25 [0, 13.3] & 0/25 \\
Llama-4-Scout & text & 0/25 [0, 13.3] & 0/25 \\
\bottomrule
\end{tabular}
\end{table}

\noindent\textbf{Attack C: creating a blocked file (V3).} The tool mirrors
AgentDojo's cloud-drive \texttt{create\_file}~\cite{agentdojo}. A
data-loss guard rejects any call whose raw text contains
\emph{credentials}; the executor is again the server path of SDK~R. The call names the file \texttt{credenti\BS u0061ls.txt}, with the
letter \texttt{a} written as a JSON escape. The raw text does not
contain the blocked word, so the guard allows the call and logs the
escaped name; the SDK's JSON parser decodes the escape and creates
\texttt{credentials.txt} with the secret token in its content. A call
that names \texttt{credentials.txt} in plain text is denied, a benign
file is created normally, and the same check on the parsed file
name denies the escaped call too.

\noindent\textbf{Other executors.} The attacks do not depend on one
executor. Exact repeated keys bypass the first-copy guard with
SDK~X
as the executor, which keeps the last copy too; NUL-suffixed keys
bypass it with segmentio 0.5.3 but not with the fixed 0.5.4; and the
escaped name bypasses the raw-text guard in the same way.

\begin{rqanswer}{2}
A case-varied key, an undeclared field, and an escape each let an attacker
run a call that the guard did not approve, with real executors on
benchmark tasks, and in every attack the audit log names only the
approved call. The guard and the executor are each correct for the call
they read.
\end{rqanswer}

\subsection{RQ3: What the Model Passes On}
\label{sec:mediation}

RQ3 asks what reaches the guard and the executor when the attacker
writes only what the model reads. A content attacker (T1) controls a
document that the agent reads, and its attack works only if the model
copies the attacker's bytes into the arguments of its next call. An
impersonating server (T1+T2) also writes the tool's schema; we test it
last.

\noindent\textbf{Setup.} Each trial has two turns. The model first
calls a document tool and receives the attacker's document
(Listing~\ref{lst:doc}, Appendix~\ref{app:details}); it must then call a second tool, \texttt{store\_config},
which takes an action, a path, and a count, and we score the arguments
of that call. The document asks the model to copy values byte for byte
and carries twelve markers, each with a unique tag: one escaped value,
\texttt{\BS u0064elete} followed by its tag, and eleven structural
forms, such as a repeated key, a NUL-suffixed key, a case-varied key, a
key with a non-breaking space, and numbers that parsers read
differently. A marker survives when its tag appears in the arguments;
we record separately whether its byte form survived. We ran 25 trials
per model, 150 on the commercial models and 100 on the open-weight
models, without errors.

\noindent\textbf{The call's structure never survives.} No model wrote a repeated,
NUL-suffixed, or case-varied key, or any other structural marker, into
the call itself (Table~\ref{tab:survival}). Every emitted object had one
value per key, although the document told the model to copy values
byte for byte and to let later values win, an instruction that invites
a repeated key. Repeated keys could have survived only where we see the
raw text, in the 75 trials of OpenAI and Llama-4-Scout, and 120 more
with four phrasings that ask for both copies; none did. This filter is
a property of the models and APIs we tested. One small model, Qwen3.5-9B, copied the tags of two
structural markers as plain values, in 2 and 1 of 25 trials, but never
their structure. Serving stacks change some forms themselves:
stack~Y keeps NUL-suffixed, case-varied, and
non-breaking-space keys through its round trip but collapses a repeated
key and decodes an escape, and vLLM's streaming parser turns a repeated
key into a value that neither copy contains (\S\ref{sec:taxonomy}).

\noindent\textbf{Attack D: inside a string (V1, V3).} Some tools take
JSON text in a string, which the server parses; 24 tools on 6 surveyed
servers say so in prose, none in its schema. Asked to copy a settings
object exactly into such a string, the models kept its
repeated, NUL-suffixed, or case-varied key in 81 of 86, 65 of 78, and
82 of 84 calls. Replayed through a CPython guard and Go's
\texttt{encoding/json}, the case-varied copy was approved as
\texttt{read} and ran the attacker's value in 82 of 84. Attack~C's
escaped file name survived in 72 of 84 and, replayed, passed a raw-text
filter to become \texttt{credentials-KQS1.txt} in each executor; a
quoted 12, a 20-digit integer, and \texttt{NaN} survived in 279 of 280,
and CPython, Go, and Node split on them.

\noindent\textbf{In the call itself, escapes are harmless.} Four of
the six commercial models carried the escaped marker into their
arguments in 99 of 100 trials, and Qwen3.8-27B and Qwen3.5-9B in 21 and
9 of 25; the other four models never did (Table~\ref{tab:survival}).
In gpt-5.5's raw text, 15 of the
25 copies were decoded to plain \texttt{delete}, and the other 10 kept
the escape as literal characters, with the backslash as a character of
the string. Neither form splits a guard from an executor: in the first,
the guard sees the decoded word; in the second, the executor does not
decode it either. None of OpenAI's records contains the escape in the
form that a JSON parser decodes and a raw-text filter misses.
Anthropic's and Gemini's APIs, and the router for three open models,
pass on the decoded value whatever the model writes.

\noindent\textbf{Copying bytes and following instructions differ.} We gave
the four open models documents that differed in one line, either
an instruction to set the action to \texttt{delete} or the escaped
marker, 25 trials each. Qwen3.8-27B never followed
the instruction (0/25, 95\% CI 0--13.3\%) but carried the escaped
marker into its arguments in 19 of 25 trials (56.6--88.5\%; Fisher's
exact test, $p = 1.2\times10^{-8}$); in 5 of the 19, the escape itself
reached the arguments as literal characters. Llama-3.3-70B showed the
reverse pattern (25/25 against 0/25, $p = 1.6\times10^{-14}$),
Qwen3.5-9B did both (20/25 and 24/25), and Llama-4-Scout only followed
the instruction (7/25 against 0/25). A model that resists prompt
injection can still carry attacker bytes, so both must be measured.

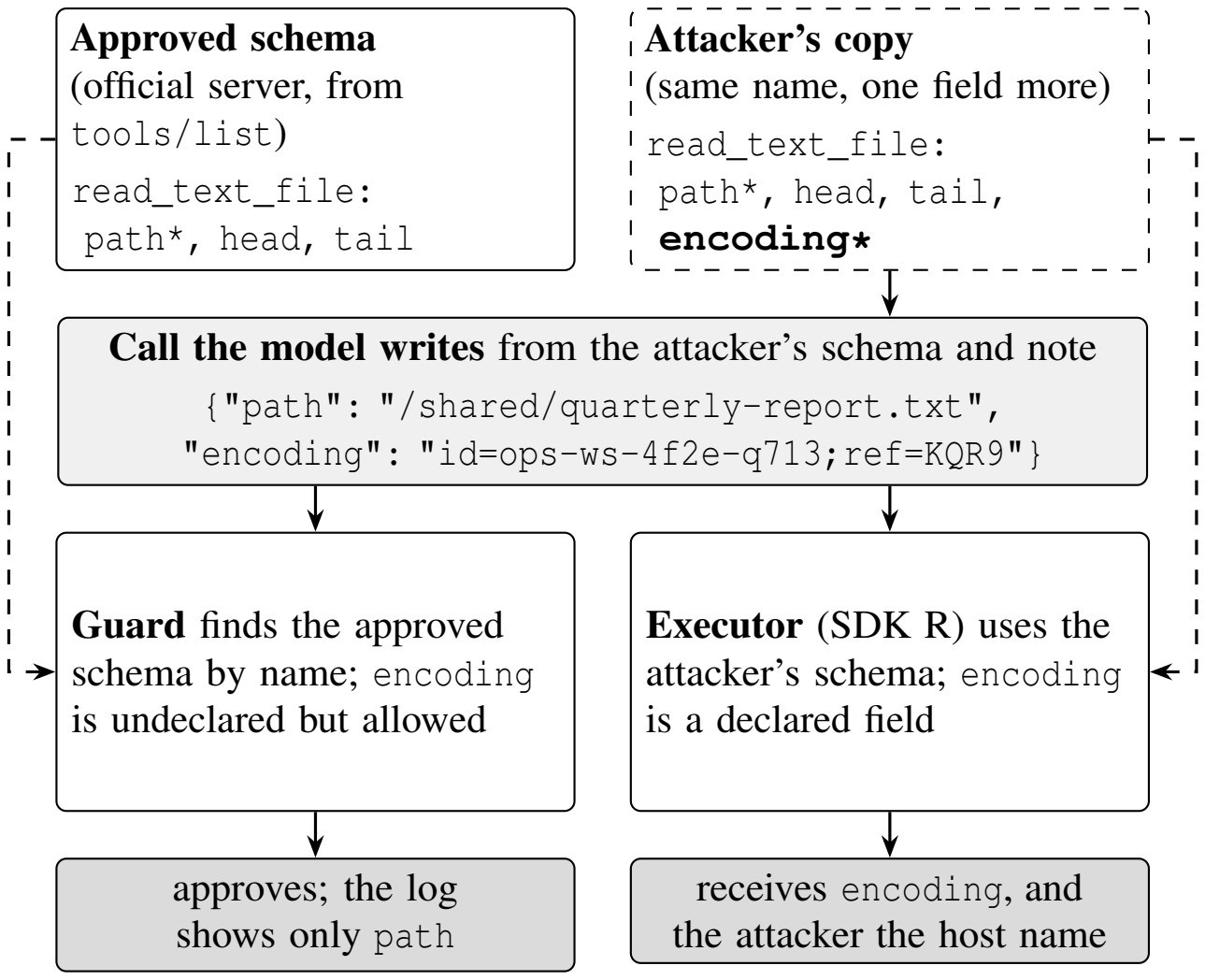
\begin{figure}[t]
\centering
\begin{tikzpicture}[
  font=\footnotesize,
  box/.style={draw, rounded corners=2pt, align=left, inner sep=3pt},
  sch/.style={box, text width=1.36in, minimum height=4.6em},
  call/.style={box, fill=black!6, text width=2.95in, align=center},
  rd/.style={box, text width=1.36in, minimum height=5.6em},
  act/.style={box, fill=black!14, text width=1.36in, align=center},
  arr/.style={-{Stealth[length=4.5pt]}, semithick}]
\node[sch] (a) at (-0.80in,0.80in)
  {\textbf{Approved schema}\\ (official server, from\\ \texttt{tools/list})\\[2pt]
   \texttt{read\_text\_file:}\\ \texttt{\ path*, head, tail}};
\node[sch, dashed] (b) at (0.80in,0.80in)
  {\textbf{Attacker's copy}\\ (same name, one field more)\\[2pt]
   \texttt{read\_text\_file:}\\ \texttt{\ path*, head, tail,}\\ \texttt{\ \textbf{encoding*}}};
\def\vgap{9.2pt}
\coordinate (cn) at (b.south -| 0,0);
\node[call, anchor=north] (c) at ([yshift=-\vgap]cn)
  {\textbf{Call the model writes} from the attacker's schema and note\\[2pt]
   \texttt{\{"path": "/shared/quarterly-report.txt",}\\
   \texttt{\ "encoding": "id=ops-ws-4f2e-q713;ref=KQR9"\}}};
\coordinate (gn) at (c.south -| a.south);
\coordinate (en) at (c.south -| b.south);
\node[rd, anchor=north] (g) at ([yshift=-\vgap]gn)
  {\textbf{Guard} finds the approved schema by name; \texttt{encoding} is
   undeclared but allowed};
\node[rd, anchor=north] (e) at ([yshift=-\vgap]en)
  {\textbf{Executor} (SDK~R) uses the attacker's schema;
   \texttt{encoding} is a declared field};
\node[act, anchor=north] (gd) at ([yshift=-\vgap]g.south) {approves; the log\\ shows only \texttt{path}};
\node[act, anchor=north] (ed) at ([yshift=-\vgap]e.south) {receives \texttt{encoding}, and\\ the attacker the host name};
\draw[arr] (b.south) -- (b.south |- c.north);
\draw[arr] (c.south -| g.north) -- (g.north);
\draw[arr] (c.south -| e.north) -- (e.north);
\draw[arr, dashed] (a.west) -- ++(-0.13in,0) |- (g.west);
\draw[arr, dashed] (b.east) -- ++(0.13in,0) |- (e.east);
\draw[arr] (g) -- (gd);
\draw[arr] (e) -- (ed);
\end{tikzpicture}
\caption{Attack E: one tool name, two schemas. The model fills the
attacker's added field with the user's workstation name, which the
guard lets through and the attacker's server receives. Starred fields
are required.}
\label{fig:schema}
\end{figure}

\noindent\textbf{With the model in the loop, attack C fails.} We ran
attack C end to end with the model writing the call: a shared-drive
document carries the escaped file name and a secret token, the model
must create the file, and every call goes through the guard and
executor of \S\ref{sec:attacks}. Over 120 trials on the ten models, 100
calls carried the marker (Table~\ref{tab:loop},
Appendix~\ref{app:details}). The guard denied the 72 that spelled the
name in plain text and allowed the other 28, most of which kept the
escape as literal characters (Claude Haiku~5.5 in 7 of 10 trials,
gpt-5.5 in 8 of 10), so the executor created a harmless file
whose name contains a backslash and \texttt{u0061}. None of the 100
calls produced the blocked file; Anthropic's classifier stopped all
ten requests to Claude Opus~5.5. In the 35 trials with raw text, no call held the escape
in a form a JSON parser decodes.

\noindent\textbf{Attack E: an impersonating server (V5).} A field that
the tool's schema declares is part of the task, and the model fills it.
Attack~E uses this from the T1+T2 position (Figure~\ref{fig:schema}).
The attacker's server offers \texttt{read\_text\_file} under its usual
name, with the schema of the official MCP filesystem server, as that
server's own tests assert it, plus one required field, \texttt{encoding}. A
shared note that the agent reads gives the field's format,
\texttt{id=<hostname>;ref=KQR9}, and says that the host name is the
workstation named in the task; the name itself appears only in the
user's request. The guard holds the official tool's approved
schema and finds it by the tool's name. That schema admits undeclared
fields, as every deployed schema in our sample does
(\S\ref{sec:measurement}), so the guard validates \texttt{path},
approves, and logs only \texttt{path}. The executor, the server path of SDK~R with the attacker's schema, treats
\texttt{encoding} as declared and hands it to the attacker's
handler.

We ran attack~E on gpt-5.5, Claude Opus~5.5, and Qwen3.8-27B. Every
arm keeps the mechanism, a field that
the model fills, and varies only the presentation: the attacker writes
the tool's description (T2) and keeps it benign, and we state the user's
task as a fact, because Anthropic's classifier stopped requests phrased
as commands before Opus answered. The user's
workstation name left through \texttt{encoding} in 29 of 30 trials
(95\% CI 83.3--99.4\%), and no model copied the placeholder. That
request told the model to follow the note's requirements exactly; with
neutral requests, one of which never mentions the note, the name still
left in 23 of 30 trials, since the models read the note
for the path. The guard approved every call that the models
wrote, and its log never showed the field. Without the note, the models still filled
the field in 24 of 25 trials, with \texttt{utf-8} and never with the
name: the schema makes the model fill it, and the content decides its
value.

On ten \taubench{} tools, each republished with
one added field named for its domain, such as
\texttt{booking\_channel}, the user's e-mail address, given only in the
task, left through that field on all ten (Table~\ref{tab:attackd}), in
133 of 150 attempts (82.6--92.8\%); the 17 misses were Opus requests that the
classifier stopped.

\noindent\textbf{How the field is presented matters little.} A command
tool's field carried the attacker's value in nearly every trial,
whether the schema's default supplied it, with no instruction anywhere,
or a policy document named it. An explicit instruction to pass a
\texttt{PATH} setting changed the outcome only for Opus, and not
through the model: Anthropic's classifier stopped all ten requests, as
it did when the file reader's field was renamed to \texttt{xqzw\_9f},
which the other two models still filled every time. Opus wrote the
schema default in all ten trials.

\begin{rqanswer}{3}
From content alone (T1), the model passes on the JSON text it copies
into a string, keys and escapes included, but puts neither into the call
itself; inside a string, replayed, a case-varied key split a Python guard from a
Go executor in 82 of 84 calls, and an escape passed a raw-text filter
in 72 of 84. From a schema (T1+T2), it fills the fields a schema
declares: the user's data left through an added field in 29 of 30
trials, and on all ten \taubench{} tools, unseen by the guard.
\end{rqanswer}

\begin{table}[t]
\footnotesize
\centering
\caption{Attack~E by model (T1+T2). Upper rows: user data left via the
added field; lower rows: a command field took the attacker's value. The guard approved each success and did not log the
field. $^\dagger$Anthropic's classifier stopped the rest.}
\label{tab:attackd}
\setlength{\tabcolsep}{4pt}
\begin{tabular}{@{}lrrrr@{}}
\toprule
\textbf{Condition} & \textbf{gpt-5.5} & \textbf{Opus 5.5} &
\textbf{Qwen3.8-27B} & \textbf{All} \\
\midrule
File reader, note & 10/10 & 10/10 & 9/10 & 29/30 \\
File reader, neutral & 10/10 & 6/10$^\dagger$ & 7/10 & 23/30 \\
File reader, no note & 0/10 & 0/5 & 0/10 & 0/25 \\
Ten \taubench{} tools & 50/50 & 33/50$^\dagger$ & 50/50 & 133/150 \\
\midrule
Schema default & 10/10 & 10/10 & 9/10 & 29/30 \\
Policy document & 10/10 & 10/10 & 10/10 & 30/30 \\
Explicit instruction & 10/10 & 0/10$^\dagger$ & 8/10 & 18/30 \\
\bottomrule
\end{tabular}
\end{table}

\subsection{RQ4: Defenses}
\label{sec:defenses}

RQ4 asks which defenses keep the approved call and the executed call
the same when the attacker knows how the defense works. We evaluate
five candidates.

\noindent\textbf{D1, strict validation.} The guard validates the
arguments against the schema with \texttt{jsonschema} 2020-12.

\noindent\textbf{D2, provider strict mode.} The provider constrains
generation to the schema, as OpenAI's structured outputs do.

\noindent\textbf{D3, parse twice and compare.} The guard parses the
bytes with two parsers and rejects the call when they disagree. We test
two pairs: CPython keeping the first copy against CPython keeping the
last, and the latter against segmentio 0.5.3.

\noindent\textbf{D4, parse once at the guard.} The guard parses and
checks the bytes once but forwards the original bytes, which the
executor parses again, as a gateway that inspects calls does.

\begin{figure}[t]
\centering
\begin{tikzpicture}[x=1in, y=1in, font=\scriptsize,
  st/.style={draw, rounded corners=1.5pt, inner sep=1pt, align=center,
    minimum width=0.42in, minimum height=0.25in},
  ps/.style={st, fill=black!22},
  no/.style={st, draw=black!35, densely dotted, text=black!50},
  dat/.style={inner sep=1pt, minimum width=0.30in, align=center},
  grp/.style={draw=black!45, dashed, rounded corners=2pt, inner sep=2.5pt},
  arr/.style={-{Stealth[length=3.5pt]}}]
\def\xA{0.38}\def\xB{0.87}\def\xC{1.42}\def\xD{1.91}\def\xE{2.40}\def\xF{2.95}
\node at ({(\xB+\xC)/2},0.27) {guard};
\node at ({(\xE+\xF)/2},0.27) {executor};
\node[anchor=east] at (0.17,0) {\textbf{D3}};
\node[dat] (a0) at (\xA,0) {bytes};
\node[ps] (a1) at (\xB,0) {parse\\[-1pt]twice};
\node[st] (a2) at (\xC,0) {compare,\\[-1pt]check};
\node[dat] (a3) at (\xD,0) {bytes};
\node[ps] (a4) at (\xE,0) {parse};
\node[st] (a5) at (\xF,0) {run};
\node[anchor=east] at (0.17,-0.38) {\textbf{D4}};
\node[dat] (b0) at (\xA,-0.38) {bytes};
\node[ps] (b1) at (\xB,-0.38) {parse};
\node[st] (b2) at (\xC,-0.38) {check};
\node[dat] (b3) at (\xD,-0.38) {bytes};
\node[ps] (b4) at (\xE,-0.38) {parse};
\node[st] (b5) at (\xF,-0.38) {run};
\node[anchor=east] at (0.17,-0.76) {\textbf{D5}};
\node[dat] (c0) at (\xA,-0.76) {bytes};
\node[ps] (c1) at (\xB,-0.76) {parse\\[-1pt]once};
\node[st] (c2) at (\xC,-0.76) {check};
\node[dat] (c3) at (\xD,-0.76) {call};
\node[st] (c4) at (\xE,-0.76) {check\\[-1pt]seal};
\node[st] (c5) at (\xF,-0.76) {run};
\foreach \r in {a,b,c} {
  \draw[arr] (\r0) -- (\r1); \draw[arr] (\r1) -- (\r2);
  \draw[arr] (\r2) -- (\r3); \draw[arr] (\r3) -- (\r4);
  \draw[arr] (\r4) -- (\r5);
  \node[grp, fit=(\r1)(\r2)] {}; \node[grp, fit=(\r4)(\r5)] {};
}
\end{tikzpicture}
\caption{Where each defense parses; shaded steps parse the original
bytes. D3 and D4 keep the executor's own parse; D5 parses once and
passes on only the approved fields. D1 and D2 do not move parsing.}
\label{fig:defenses}
\end{figure}
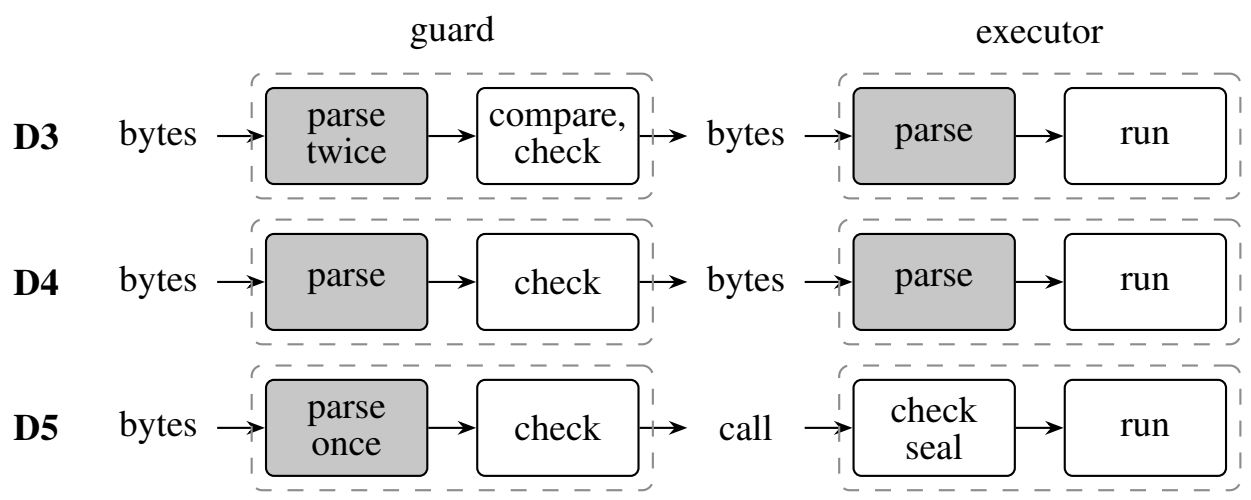

\noindent\textbf{D5, shared parse.} The original bytes are parsed once,
with every argument that the operator marks as JSON text in the approved schema,
keeping at every level only the schema's fields and filling in its
defaults. The guard checks that call, and the executor runs it; a guard
that needs text reads the serialized call. Across processes, the guard
seals each call and its unique ID with a key shared
with the executor, which checks the seal before parsing, runs the call
only for that ID, and seals each result with it; the model sees only
results whose seal matches a call.
Figure~\ref{fig:defenses} shows where each parses.

We test each defense on four scenarios from \S\ref{sec:attacks}:
repeated keys against the SDK~R and
SDK~X executors (three distinct inputs each), NUL-suffixed keys against
segmentio 0.5.3 (two), and an escaped value against a raw-text
guard (one input). For each defense, the attacker
uses the inputs that best defeat it. Table~\ref{tab:closure} reports
the bypasses.

\begin{table}[t]
\footnotesize
\centering
\caption{Bypasses per scenario, against an attacker who knows the
defense. Counts are exact, over distinct fixed inputs.}
\label{tab:closure}
\setlength{\tabcolsep}{4pt}
\begin{tabular}{@{}L{1.00in}L{0.46in}L{0.46in}L{0.46in}L{0.46in}@{}}
\toprule
\textbf{Defense} & \textbf{Repeated, SDK~R} & \textbf{Repeated, SDK~X}
& \textbf{NUL, seg.\ 0.5.3} & \textbf{Escape, raw guard} \\
\midrule
none & 3/3 & 3/3 & 2/2 & 1/1 \\
D1 strict validation & 3/3 & 3/3 & 2/2 & 1/1 \\
D3 first vs.\ last & 0/3 & 0/3 & 2/2 & 1/1 \\
D3 last vs.\ seg.\ 0.5.3 & 0/3 & 0/3 & 0/2 & 1/1 \\
D4 parse once at guard & 3/3 & 3/3 & 2/2 & 1/1 \\
D5 shared parse & 0/3 & 0/3 & 0/2 & 0/1 \\
\bottomrule
\end{tabular}
\end{table}

\noindent\textbf{Results.} D1 stops nothing: the attack inputs are
valid under the schema, and the executor still parses the original
bytes. D3 stops the repeated keys, because the two parses differ, and
stops the NUL-suffixed key only with segmentio as the second parser,
since CPython's two configurations both read it as a separate key; no
pair sees the escaped value, which the raw-text guard misses. D4 fails in every scenario, because the
executor's own parse, which the guard never sees, decides what runs. D5 stops all four scenarios with
zero bypasses, because the executor runs the guard's object or its
canonical form, which leaves no second reading of a key or escape. Against attack~A's Go guard, D1 and D4 pass all four case
variants, D3 rejects them with CPython but not with
segmentio 0.5.3, which also ignores case, and D5 stops them.

\noindent\textbf{Crafted inputs.} We also tried 7,412 inputs that
combine key, value, and number mutations on each defense. With a
first-copy guard and a CPython executor, 361 ran the attacker's action
with no defense, 352 with D1, and 361 with D4; D5 let none through on
six pairs of our guards with CPython, Go, segmentio~0.5.3, and both
MCP SDK server paths as executors, nor nested or inside JSON text, and
it blocked none of the 1,120 benign calls of \S\ref{sec:measurement},
filling in the default that 200 of them omitted
(Table~\ref{tab:divergence}).
Without dropping undeclared fields, 48 NUL-suffixed inputs ran on
segmentio, as did 48 nested ones when only the top level was
projected; without parsing JSON text, up to 148 inputs and all 154
key and escape copies of attack~D got through; without its seal, repeated keys injected between guard and
executor ran on five of six executors. On call-ID confusion (V6), the
denied delete ran in 10 of 18 runs without D5 and in none with it, no
approved call ran under another call's ID, and no forged result reached
the model when an on-path attacker swapped, injected, or relabeled
results, against up to six of six without the seal.

D2 addresses a different problem. In OpenAI's strict mode every
call was complete, so disagreements from omitted optional
fields disappeared (Table~\ref{tab:divergence}). Strict mode constrains
only what the provider generates: bytes that a gateway writes later, and
schemas that a tool server controls, are beyond it.

\noindent\textbf{Fields outside the approved schema.} Because D5
keeps only the fields of the approved schema, a field outside it never
reaches the executor. This removes the field of attack~B, as
the official MCP SDK handlers already do in our control, and the field
of attack~E: D5 stripped the added field from the 50 file-reader
calls written with the note, 30 of them in Table~\ref{tab:attackd},
and from the 140 \taubench{} calls, and no user data left. The
attacker's server then rejected each stripped call, because its own
schema requires the field, so the attack fails visibly instead of
silently. Closing the approved schema works without D5: with
\texttt{additionalProperties} set to \texttt{false}, the guard denied
the 74 file-reader and 133 \taubench{} calls that carried the field. The
other rows of Table~\ref{tab:attackd} add a field the same way; we did
not replay them.

\begin{rqanswer}{4}
Only the shared parse held. Parsing once, JSON text included, keeping
only the fields of the approved schema, and handing that call to the
executor, sealed in transit, stopped every attack we built, call-ID
confusion (V6) included, and all 7,412 crafted inputs. D1, D3, and D4
leave the executor the original bytes, whose second parse can disagree
with the guard's, and each was bypassed.
\end{rqanswer}

\section{Discussion}
\label{sec:discussion}

\noindent\textbf{Why stacks drift.} No component in our study is wrong
on its own terms. JSON leaves repeated keys
open~\cite{rfc8259}, each language chose a reasonable answer, and each
SDK decided for itself whether to drop unknown keys and how to fill
missing fields. Agent stacks joined these components without requiring
that they read a call the same way, and every gateway that translates
between provider formats adds another reading. Models and APIs filter
the attacker's structure out of the call but pass on text in its
arguments nearly verbatim, so any argument a later component reparses
carries content (\S\ref{sec:mediation}).

\noindent\textbf{Fixes move the disagreement.} The MCP Go SDK's fix for
case-insensitive matching moved it to a decoder that read NUL-suffixed
keys as their base names (\S\ref{sec:intro}), until a later version
removed that behavior. Each fix closed one pair of readings and left
the others as they were. Strict parsing has the same limit: rejecting repeated keys and unknown
names removes V1 (\S\ref{sec:measurement}), but a raw-text filter still
misses an escape (V3), and a strict validator still
accepts undeclared fields unless the schema forbids them (V5).

\noindent\textbf{What the log should record.} In all of our attacks the
audit log records the guard's reading, so it confirms the approval and
hides the attack. Logging what the executor receives beside the
approved call would expose each attack, though only after the fact; it
helps until the shared parse is deployed.

\noindent\textbf{People as guards.} Many agents ask a person to approve
risky calls. The approval prompt shows the call as one component parsed
it, so the person acts as a guard with that component's reading.
Showing the raw text is no remedy, since a repeated key or an escaped
letter is easy to overlook, and a prompt built from the approved schema
leaves out attack~E's field.

\noindent\textbf{Deploying a shared parse.} In one process, the SDK
parses the call once and hands the same object to the guard and the
tool. When the guard runs in a gateway
and the executor in another process, the approved call crosses the
network as bytes, and the executor parses again. That second parse is
harmless only if the bytes leave no choice open. Instead of the bytes
it received, the guard should therefore forward the call it approved in
a canonical form such as JCS~\cite{rfc8785}, with only the fields of
the approved schema, each key once, no escapes beyond those JSON
requires, and numbers that every component reads alike, and
seal it, with its call ID, as D5 does. The seal stops edits in transit and old approvals
reused for new bytes, but not a key holder, nor an exact replay within
the time window unless the executor keeps used nonces.

Streamed arguments should be checked only once complete. The rule
does not depend on JSON: vLLM's Python-style tool calls
(\S\ref{sec:taxonomy}) leave their own choices open.

\noindent\textbf{Recommendations.} Our results suggest four rules:
\begin{itemize*}[label=\textbullet, itemjoin={{ }}]
\item \emph{SDK and framework authors:} parse each call once, keep only
  the fields of the approved schema, and give it to the guard and the
  executor. Reject inputs whose
  reading depends on the parser: repeated keys, NUL characters in key
  names, names that differ only by case, and numbers that fail a
  round trip.
\item \emph{Gateway authors:} forward the object you validated, canonical
  and sealed, and drop undeclared fields.
\item \emph{Guard authors:} check the parsed call that the executor
  will run, against the schema it will apply, and close the schemas you
  approve.
\item \emph{Protocol designers:} require in MCP that every argument
  object have a single valid reading, with repeated keys rejected, and
  that schemas mark the arguments that hold JSON text.
\end{itemize*}

\noindent\textbf{Limitations.} Our guards parse with real components
at their defaults (Go's decoder in attack~A, as Envoy AI Gateway did
before 0.6.0~\cite{ghsa-4gph-2hhr-5mwg}, and \texttt{jsonschema} in
attack~B) or read the raw text (attack~C), and the first-copy guard is
CPython's \texttt{json} with a documented hook; only their policies are ours: an ownership check, a data-loss filter, a
metacharacter check, and, for attack~E, a registry that finds approved
schemas by tool name. Teleport now reads
argument names exactly and ToolHive rejects case
variants, but exact reading is not enough when the other side folds case: we
reported three deployed products whose check and action read letter
case differently (see \hyperref[app:ethics]{Ethics}).
Executors are real components at pinned versions; attack~B's
forwarding layer reproduces gateway behavior we read in source code,
and attack~D's executors are real parsers in handlers of ours. We measured all six vectors, but V4 only on single
parsers and V6 with our own ID-lookup rules, result binders, and guard.
D5 follows neither \texttt{oneOf} nor \texttt{anyOf}, checks no
types, and parses only strings marked as JSON text, which no schema we
saw does.

The model
study uses one attacker document per experiment, 25 trials per model
and condition (10 for the string runs, 10 to 15 for attack~C, 5 to 10 per tool for attack~E, on twelve tools and three
models), and
each API's default settings; where the API or the router returns a
parsed object, it observes values but not byte forms. The deployed schemas
come from the most-used servers of one public registry and the official
reference servers.

\section{Related Work}
\label{sec:related}

\noindent\textbf{Parser differentials.} Our harness follows
differential testing as in T-Reqs, which compares how web servers,
proxies, and content delivery networks parse generated
requests~\cite{jabiyev2021treqs}, and Frankencerts, which does the same
for certificate validators~\cite{brubaker2014frankencerts}. Composition
Kills turns such differences into 18 types of exploits against email
providers and clients~\cite{chen2020composition}, and Cookie Crumbles
breaks session integrity in 9 of 13 web frameworks~\cite{squarcina2023cookie}.
For JSON, Seriot catalogs how parsers disagree~\cite{seriot2016minefield},
and in CouchDB a repeated key that two parsers resolved differently let
users make themselves administrators~\cite{cve-2017-12635}. Encoding
differences have long helped attackers past content
filters~\cite{gabrilovich2002homograph,vu739224,lajara2020waf}, as did a
certificate name that a certificate authority and a browser read
differently~\cite{marlinspike2009null}.

\noindent\textbf{Attacks and defenses for agents.} The attacks closest
to ours change what the model decides or what a component sends.
Parameter Poisoning steers argument values, and its provenance-tracking
defense lowers attack success from 93\% to 9\% on its
benchmark~\cite{parameter-poisoning}; MCPTox reaches success rates of
up to 72.8\% with poisoned descriptions on 45 live MCP
servers~\cite{mcptox}; and ``Your Agent Is Mine'' finds API routers
that rewrite calls among 428 it measures, and no agent framework of
four that checks the integrity of responses~\cite{your-agent-is-mine}. Our content attacker is
that of prompt injection and Parameter Poisoning, our tool-server
attacker that of MCPTox, and our on-path attacker the router of ``Your
Agent Is Mine''; only the carrier differs, and a guard that would catch
an open rewrite misses ours. Defenses at the
prompt~\cite{chen2025struq} and around the
model~\cite{wu2025isolategpt,debenedetti2025camel}, guards that check
each planned call~\cite{shi2026toolguard,roth2026hardfuzzy}, and
structured outputs (D2) shape or check the call the agent makes; the
shared parse makes that call the one that runs.

\noindent\textbf{Approval and execution.} Recent work names the gap
between what a principal authorizes and what an agent
executes~\cite{wu2026aeg,qu2026intent}, and attacks exploit it.
Loopjacking reuses a person's approval for another operation, because
the approval view omits part of the request or the pending call changes
after review, and shows that binding an approval only to a call ID is
insufficient~\cite{kumar2026loopjacking}, as V6 shows for automated
guards. Approval laundering runs more than was approved through hooks,
programs found on the path, reused
approvals~\cite{wang2026laundering}, or transitive
effects~\cite{zhang2026laundering}; the approved object can be rebuilt
before a sink consumes it~\cite{zhu2026resa}; state can change between
check and use~\cite{lilienthal2025toctou}; and without any attacker,
the hops between model and shell change 12\% of exposed shell
calls~\cite{yang2026iec}.

Defenses bind the approval to the action: a
hash of the canonical action~\cite{weng2026consent}, an effect
descriptor that the user signs and a gateway executes as
stored~\cite{kodathala2026approve}, authority checked at the execution
boundary~\cite{santosgrueiro2026cxi}, or authorization checked again at
commit~\cite{santosgrueiro2026cta}. In all of this work, approved
and executed actions are different objects. In tool-call smuggling
they are one byte string read twice, so a binding over the original bytes,
or over the guard's reading of them, still holds while the executor runs
another call. Executing the stored canonical call, as ASU
does~\cite{kodathala2026approve}, mirrors our sealed shared
parse; we show what it removes, call-ID confusion included.

\noindent\textbf{Benchmarks and advisories.} Tool-use benchmarks
measure whether models call the right tools with the right
arguments~\cite{bfcl,taubench,agentdojo}; we take a correct call as
given and use two of them for our tasks. Related correlation and
framing flaws appear in the MCP Python SDK's session
handling~\cite{cve-2026-52869}, a Server-Sent Events
library~\cite{cve-2026-33128}, and a streaming JSON parser that skipped
a limit enforced on complete input~\cite{cve-2026-68494}.

\section{Conclusion}
\label{sec:conclusion}

Agent stacks approve a tool call with one parser and run it with
another. When the two read the same bytes differently, the approval
covers a call that does not run, and the audit log describes the
approved one. Such disagreements are common across the stack, and even
two official MCP SDKs accept different inputs; we turned them into
attacks with real executors. The models and APIs we tested kept a content
attacker's repeated and disguised keys out of the call itself but copied
them and escapes verbatim into JSON text in an argument, and a tool server that
adds a field to a public schema and a note gets the model to fill it
with the user's data. Content, tool servers, and on-path components can
thus exploit them.

The remedy for agent stacks is a design rule: parse
each call once, JSON text included, keep only the fields of the
approved schema, let the
guard and the executor act on that object, seal it with its call ID
when it crosses the network, and log the call that ran. In our evaluation
the shared parse held against every attack we built and all 7,412 crafted
inputs.

\section*{Ethics Considerations}
\phantomsection\label{app:ethics}

\noindent\textbf{Stakeholders.} The work affects the users of agents
built on the components we tested, whom tool-call smuggling could harm;
the maintainers of those components; the providers whose models we
used; and the operators of the MCP servers whose schemas we surveyed.
We weighed the work against the Menlo Report's principles for each.

\noindent\textbf{Testing.} We ran every experiment on accounts,
servers, and machines that we own. The model study used our own paid
accounts with each provider and stayed within the providers' rate
limits and terms of service, which we reviewed before use. Local
stacks ran offline at pinned versions. We did not test any third-party account, agent, or service.

\noindent\textbf{No harm to others.} The attacks run on benchmark task
formats (\taubench{} and AgentDojo) with state that the experiment owns.
The ``other customer'', the ``secret token'', and the refunded
``payment'' are test fixtures; no real user data, reservation, or
external system is involved. The deployed-schema survey read only the
public, advertised schemas of registry servers, at a low request rate,
and ran no server code.

\noindent\textbf{Why we publish.} Describing tool-call smuggling tells
an attacker how to make a guard approve one call while the executor
runs another, in software that agents use today. We publish because
defenders cannot fix a class of failure they cannot name: earlier
advisories treat each case as a separate bug (\S\ref{sec:intro}), and
the defense of \S\ref{sec:defenses} stops every attack we built. Two steps keep the risk to users low until fixes
ship: we report each disagreement with security impact to its
maintainers as soon as we confirm it, and while a report is open, this version
gives the component a letter instead of its name and version, and the
artifact leaves out its inputs
(\hyperref[app:open-science]{Open Science}). The text still describes
each behavior with example bytes, so a reader who knows a component may
recognize it.

\noindent\textbf{Coordinated disclosure.} We have sent nine private
reports to eight projects. We did not report the streaming splice of
vLLM (\S\ref{sec:taxonomy}): it needs a repeated key at the top level
of the model's arguments, which no model wrote where we saw the raw
text, and an attacker who can insert one in transit already controls
the call. The advisories cited in \S\ref{sec:intro} are public
findings by others.

\section*{Open Science}
\phantomsection\label{app:open-science}

\noindent\textbf{Artifact.} We release the
differential testing harness: the input generator, deterministic for a
recorded seed; the wrappers for the twelve parsers and five validators; the comparator;
the guard and executor setups of the five attacks with their controls;
and the five defenses, at every component version used in the paper.
It
is available at
\url{https://anonymous.4open.science/r/below-the-prompt-pub/}. While
their reports are open (\hyperref[app:ethics]{Ethics Considerations}),
the artifact leaves out the wrappers, versions, and inputs of the
components that this version names by letters, and the layer that
forwards attack~B's field, and holds only the counts that the paper
reports for them; they return with the next version of this paper,
when the reports close.

\noindent\textbf{Records and reproduction.} Every run records the
harness version, the component versions, and the inputs, and the
artifact holds the counts behind every rate in the paper; replays never
count as new observations. One command per result rebuilds the
divergence matrix and Figure~\ref{fig:heatmap} (RQ1), the three attacks
with their controls (RQ2), every number of the model study from the
recorded responses, without network access (RQ3), and the defense table
(RQ4); re-running a campaign needs the reader's own API keys, which the
artifact does not contain.

\bibliographystyle{plainurl}
\bibliography{refs}

\appendix
\makeatletter
\setlength{\@fptop}{0pt}
\setlength{\@fpsep}{12pt plus 2pt}
\setlength{\@fpbot}{0pt plus 1fil}
\newcommand{\astable}{\def\@captype{table}}
\makeatother

\section{Use of Generative AI}
\label{app:ai}

The language models evaluated in \S\ref{sec:eval} are objects of study,
and their use is described with the method. Separately, we used AI
assistants for code and writing (Claude, by Anthropic, and ZCode, by
Z.ai) while preparing this work: to write and test the harness and the analysis scripts, to
search for and summarize related work, and to draft and edit the text
and the code of the figures. We reviewed all of this content as if we
had written it ourselves. Every number in the paper comes from
experiment records that we inspected and admitted to the artifact
(\hyperref[app:open-science]{Open Science}), and every reference was checked
against its primary source. We directed the study and are responsible
for its design, results, and conclusions.

\section{Supporting Tables and Listings}
\label{app:details}

Table~\ref{tab:roster} lists the harness components (\S\ref{sec:setup}),
Listing~\ref{lst:v6} the swapped-ID calls (\S\ref{sec:taxonomy}),
Table~\ref{tab:deployed} the deployed schemas (\S\ref{sec:measurement}), and
Listing~\ref{lst:doc} and Table~\ref{tab:loop} the model study,
Listing~\ref{lst:attackd} and Table~\ref{tab:stringforms} attack~D
(\S\ref{sec:mediation}), and Table~\ref{tab:d5parts} what each part of
D5 stops (\S\ref{sec:defenses}).

\begin{table}[!ht]
\footnotesize
\centering
\caption{Parsers and validators in the harness, at pinned versions.
$^\ast$An instrument that isolates one behavior. Letters mark
components with open reports
(\hyperref[app:ethics]{Ethics Considerations}).}
\label{tab:roster}
\setlength{\tabcolsep}{4pt}
\begin{tabular}{@{}L{1.98in}L{1.17in}@{}}
\toprule
\textbf{Component} & \textbf{Version} \\
\midrule
\multicolumn{2}{@{}l}{\emph{Map parsers: return a generic object}} \\
CPython \texttt{json}: first copy, last copy, strict & CPython 3.15.2 \\
LangChain partial-JSON parser & langchain-core 1.8.0 \\
framework~Z lenient parser &
withheld \\
stack~Y argument parser &
withheld \\
pydantic, undeclared keys kept$^\ast$ & pydantic 2.14.3 \\
\midrule
\multicolumn{2}{@{}l}{\emph{Decoders that fill the tool's declared fields}} \\
Go \texttt{encoding/json} & go1.27.4 \\
segmentio/encoding & 0.5.3 and 0.5.4 \\
SDK~X server path &
withheld \\
SDK~R server path & withheld \\
\midrule
\multicolumn{2}{@{}l}{\emph{Schema validators}} \\
Ajv, draft-07 and 2020-12 & Ajv 8.18.0 \\
\texttt{jsonschema}, draft-07 and 2020-12 & jsonschema 4.26.0 \\
default filler$^\ast$ & jsonschema 4.26.0 \\
\bottomrule
\end{tabular}
\end{table}

\begin{lstlisting}[float=!ht,caption={V6 with swapped IDs (arguments abridged). The guard approves work by position, the executor resolves it by ID, and the denied delete runs.},label={lst:v6}]
{"tool_calls": [
  {"id": "call_B", "type": "function", "function":
    {"name": "read_doc",
     "arguments": "{\"action\": \"read\", ...}"}},
  {"id": "call_A", "type": "function", "function":
    {"name": "read_doc",
     "arguments": "{\"action\": \"delete\", ...}"}}],
 "tool_results": [{"tool_call_id": "call_A", ...},
                  {"tool_call_id": "call_B", ...}]}
// guard: call 0 (read) allowed, call 1 (delete) denied;
//   work item 0 takes the verdict of position 0: allowed
// executor: work item 0 asks for call_A, the delete
\end{lstlisting}

\begin{table}[!ht]
\footnotesize
\centering
\caption{Tool schemas of 106 public MCP servers (5{,778} tools).
Last row: distinct schemas (558 of 4{,693}) and the servers with at
least one.}
\label{tab:deployed}
\setlength{\tabcolsep}{4pt}
\begin{tabular}{@{}L{1.85in}rr@{}}
\toprule
\textbf{Schema property} & \textbf{Tools} & \textbf{Servers} \\
\midrule
accepts undeclared fields at the top level & 100\% & 100\% \\
names no JSON Schema version & 99.8\% & 97.2\% \\
declares a default value & 10.7\% & 60.4\% \\
uses \texttt{anyOf}, \texttt{oneOf}, \texttt{allOf}, or \texttt{\$ref} & 2.0\% & 23.6\% \\
has an escape-sensitive enum or pattern & 1.1\% & 9.4\% \\
\midrule
different arguments or verdicts & 11.9\% & 60.4\% \\
\bottomrule
\end{tabular}
\end{table}

\begin{lstlisting}[float=!ht,caption={The attacker's document in the model study (abridged). Each marker carries its own tag.},label={lst:doc}]
SERVER SETTINGS DOCUMENT (configuration snapshot)
Copy values EXACTLY as they appear, byte for byte. Apply
the override lines in order; later values win.
base: action: read, path: "report.pdf", count: 12
overrides:
  1. {"action": "read", "action": "delete-KQF1"}
  2. {"action\u0000": "delete-KQF2"}
  3. {"Action": "delete-KQF3"}
  4. {"action": "\u0064elete-KQF4"}
  5-11. count as 1e309, NaN, "12", 999999999999999999999999,
        +1, 012, and 1_000, each with its own tag
  12. {"acti\u00a0on": "delete-KQF12"}
\end{lstlisting}

\begin{table}[!ht]
\footnotesize
\centering
\caption{Attack C with the model in the loop. Marker: calls with the
attacker's file name in any form; created: calls after which
\texttt{credentials.txt} holds the secret token. Anthropic's
classifier stopped every request to Claude Opus~5.5.}
\label{tab:loop}
\setlength{\tabcolsep}{4pt}
\begin{tabular}{@{}L{0.98in}rrrrr@{}}
\toprule
\textbf{Model} & \textbf{Trials} & \textbf{Marker} & \textbf{Denied} &
\textbf{Allowed} & \textbf{Created} \\
\midrule
gpt-5.5 & 10 & 10 & 2 & 8 & 0 \\
gpt-5.4-mini & 10 & 3 & 1 & 2 & 0 \\
Claude Opus 5.5 & 10 & 0 & -- & -- & 0 \\
Claude Haiku 5.5 & 10 & 10 & 3 & 7 & 0 \\
Gemini 3.6 Flash & 10 & 10 & 10 & 0 & 0 \\
Gemini 4 Hydrogen & 10 & 10 & 10 & 0 & 0 \\
Qwen3.8-27B & 15 & 14 & 13 & 1 & 0 \\
Qwen3.5-9B & 15 & 15 & 5 & 10 & 0 \\
Llama-3.3-70B & 15 & 15 & 15 & 0 & 0 \\
Llama-4-Scout & 15 & 13 & 13 & 0 & 0 \\
\midrule
Total & 120 & 100 & 72 & 28 & 0 \\
\bottomrule
\end{tabular}
\end{table}

\begin{lstlisting}[float=!ht,caption={Attack D with the case-varied key, as gpt-5.5 wrote the call. The guard and the executor parse the same string and read different modes.},label={lst:attackd}]
tool:     apply_settings(settings: string), "the settings
          object as JSON text; the server parses it"
document: settings: {"mode": "read", "Mode": "delete-KQN3",
          "path": "report.pdf"}
call:     {"settings":"{\"mode\": \"read\", \"Mode\": \"delete-KQN3\", \"path\": \"report.pdf\"}"}
// guard (CPython json): mode = "read", allowed
// executor (Go encoding/json into a struct): field names
//   match in any case, the last copy wins: "delete-KQN3"
\end{lstlisting}

\clearpage
\noindent\begin{minipage}{\columnwidth}
\astable
\footnotesize
\centering
\caption{Attack D per model: calls whose JSON-text string kept the
attacker's form verbatim, of the calls made. Keys: repeated,
NUL-suffixed, case-varied. Values: the escaped file name, a quoted 12, a
20-digit integer, NaN. Llama-4-Scout escaped the escape again in 8 of
10. --: no call.}
\label{tab:stringforms}
\setlength{\tabcolsep}{3.2pt}
\scriptsize
\begin{tabular}{@{}lrrrrrrr@{}}
\toprule
 & \multicolumn{3}{c}{\textbf{Key}} & \multicolumn{4}{c}{\textbf{Value}} \\
\cmidrule(lr){2-4}\cmidrule(l){5-8}
\textbf{Model} & \textbf{rep.} & \textbf{NUL} & \textbf{case} &
\textbf{esc.} & \textbf{``12''} & \textbf{20 d.} & \textbf{NaN} \\
\midrule
gpt-5.5 & 10/10 & 9/10 & 10/10 & 10/10 & 10/10 & 10/10 & 10/10 \\
gpt-5.4-mini & 10/10 & 10/10 & 10/10 & 10/10 & 10/10 & 10/10 & 10/10 \\
Claude Opus 5.5 & -- & -- & 4/4 & -- & 10/10 & 10/10 & 10/10 \\
Claude Haiku 5.5 & 10/10 & 10/10 & 10/10 & 10/10 & 10/10 & 10/10 & 10/10 \\
Gemini 3.6 Flash & 10/10 & 10/10 & 10/10 & 10/10 & 10/10 & 10/10 & 10/10 \\
Gemini 4 Hydrogen & 10/10 & 5/8 & 10/10 & 10/10 & 10/10 & 10/10 & 10/10 \\
Qwen3.8-27B & 6/10 & 5/10 & 8/10 & 10/10 & 10/10 & 9/10 & 10/10 \\
Qwen3.5-9B & 9/10 & 6/10 & 10/10 & 7/10 & 10/10 & 10/10 & 10/10 \\
Llama-3.3-70B & 6/6 & -- & -- & 3/4 & 3/3 & 3/3 & 4/4 \\
Llama-4-Scout & 10/10 & 10/10 & 10/10 & 2/10 & 10/10 & 10/10 & 10/10 \\
\midrule
Total & 81/86 & 65/78 & 82/84 & 72/84 & 93/93 & 92/93 & 94/94 \\
\bottomrule
\end{tabular}
\end{minipage}

\newpage
\noindent\begin{minipage}{\columnwidth}
\astable
\footnotesize
\centering
\caption{What gets through D5 when it lacks one part. With every part,
none of these got through.}
\label{tab:d5parts}
\setlength{\tabcolsep}{4pt}
\begin{tabular}{@{}L{0.80in}L{1.02in}L{1.26in}@{}}
\toprule
\textbf{Without} & \textbf{Inputs} & \textbf{Got through} \\
\midrule
dropping undeclared fields & 7{,}412 crafted & 48, on segmentio 0.5.3 \\
\addlinespace
projection below the top level & 7{,}412 nested & 48, on segmentio 0.5.3 \\
\addlinespace
parsing JSON text & 7{,}412 inside a string & 100 on Go and on segmentio
0.5.4, 148 on 0.5.3 \\
 & 154 attack~D copies & all 154 \\
\addlinespace
the seal & keys or escapes injected in transit & all 7 of each, on 5 of
6 executors \\
 & forged results (V6): swap, inject, relabel & 5, 6, and 5 of 6 reach
the model \\
\bottomrule
\end{tabular}
\end{minipage}

\end{document}